\documentclass[sigconf,nonacm]{acmart}

\usepackage{subfigure}
\usepackage{siunitx}
\usepackage{enumitem}

\usepackage{algorithm}
\usepackage[noend]{algpseudocode}
\let\ReturnInline\Return
\renewcommand{\Return}{\State\ReturnInline}
\algrenewcommand\algorithmicrequire{$\rhd$}
\algrenewcommand\algorithmicensure{$\square$}

\AtBeginDocument{%
  \providecommand\BibTeX{{%
    \normalfont B\kern-0.5em{\scshape i\kern-0.25em b}\kern-0.8em\TeX}}}

\setcopyright{acmcopyright}
\copyrightyear{2018}
\acmYear{2018}
\acmDOI{XXXXXXX.XXXXXXX}

\acmConference[Conference acronym 'XX]{Make sure to enter the correct
  conference title from your rights confirmation emai}{June 03--05,
  2018}{Woodstock, NY}

\newcommand{\ignore}[1]{}
\newcommand{\ok}[1]{}

\begin{document}

\title[A Fast Parallel Approach for Neighborhood-based Link Prediction by Disregarding Large Hubs]{A Fast Parallel Approach for Neighborhood-based \\Link Prediction by Disregarding Large Hubs}


\author{Subhajit Sahu}
\email{subhajit.sahu@research.iiit.ac.in}
\affiliation{%
  \institution{IIIT Hyderabad}
  \streetaddress{Professor CR Rao Rd, Gachibowli}
  \city{Hyderabad}
  \state{Telangana}
  \country{India}
  \postcode{500032}
}


\settopmatter{printfolios=true}

\begin{abstract}
Link prediction can help rectify inaccuracies in various graph algorithms, stemming from unaccounted-for or overlooked links within networks. However, many existing works use a baseline approach, which incurs unnecessary computational costs due to its high time complexity. Further, many studies focus on smaller graphs, which can lead to misleading conclusions. Here, we study the prediction of links using neighborhood-based similarity measures on large graphs. In particular, we improve upon the baseline approach (IBase), and propose a heuristic approach that additionally disregards large hubs (DLH), based on the idea that high-degree nodes contribute little similarity among their neighbors. On a server equipped with dual 16-core Intel Xeon Gold 6226R processors, DLH is on average $1019\times$ faster than IBase, especially on web graphs and social networks, while maintaining similar prediction accuracy. Notably, DLH achieves a link prediction rate of $38.1M$ edges/s and improves performance by\ignore{at a rate of} $1.6\times$ for every doubling of threads.
\end{abstract}

\keywords{Parallel Link prediction, Local/Neighborhood-based}


\maketitle

\section{Introduction}
\label{sec:introduction}
Most real-world networks are incomplete \cite{kim2011network, wang2014link}. These networks lie somewhere in the range of a deterministic and a purely random structure, and are thus partially predictable \cite{lu2015toward}. Link prediction is the problem of identifying potentially missing/undiscovered connections in such networks \cite{marchette2008predicting, kim2011network}, or even forecasting future connections \cite{bringmann2010learning, juszczyszyn2011link}, by examining the current network structure\ignore{\cite{zhou2021progresses}}. This is useful in various applications, such as recommending items for online purchase \cite{akcora2011network}, helping people to find potential collaborators \cite{mori2012machine, tang2012cross},\ignore{predicting co-authorships in academic research networks \cite{pavlov2007finding, wohlfarth2008semantic}, identifying abnormal communications \cite{huang2009time}} assessing the trustworthiness of individuals \cite{alnumay2019trust}, uncovering criminal activities and individuals \cite{berlusconi2016link, lim2019hidden}\ignore{, detecting anomalies \cite{huang2006link}}, and predicting new protein-protein interactions or generating hypotheses \cite{cannistraci2013link, nasiri2021novel}.

Similarity measures are frequently employed to predict the likelihood of missing or future links between unconnected nodes in a network \cite{wang2014link, arrar2023comprehensive}. The principle is straightforward: higher similarity indicates a greater likelihood of connection \cite{wang2014link}. The choice of metric depends on the network's characteristics, with no single metric dominating across different datasets \cite{arrar2023comprehensive, zhou2021progresses}. Local / neighborhood-based similarity metrics such as Common Neighbors\ignore{\cite{newman2001clustering}}, Jaccard Coefficient\ignore{\cite{jaccard1901etude}}, S{\o}rensen Index\ignore{\cite{sorensen1948method}}, Salton Cosine similarity\ignore{\cite{salton1973specification}}, Hub Promoted\ignore{ \cite{liben2003link}}, Hub Depressed, Leicht-Holme-Nerman, Adamic-Adar\ignore{\cite{adamic2003friends}}, and Resource Allocation\ignore{\cite{zhou2010solving}}, which are based on neighborhood information within a path distance of two, remain popular \cite{arrar2023comprehensive, wang2014link}. This is due to their simplicity, interpretability \cite{pai2019netdx, barbieri2014follow}, computational efficiency \cite{garcia2014link}, and the ability to capture underlying structural patterns.\ignore{Further, they are often combined with other metrics.}

However, many studies \cite{gatadi2023lpcd, saifi2023fast, benhidour2022approach, mumin2022efficient, rafiee2020cndp, guo2019node, yang2015new, papadimitriou2012fast, wang2019link} and network analysis software \cite{staudt2016networkit, csardi2006igraph} use a baseline approach for link prediction, computing unnecessary similarity scores for all non-connected node pairs. Further, early studies often evaluate a limited number of algorithms on small networks --- this can result in misleading conclusions \cite{zhou2021progresses, zhou2021experimental}. Further, much-existing research does not address link prediction for large networks with close to a billion edges \cite{muscoloni2022adaptive, mumin2022efficient, nasiri2021novel, xian2021towards, ghasemian2020stacking, mara2020benchmarking, wang2019link, xu2019distributed, mohan2017scalable, cui2016bounded, garcia2014link, papadimitriou2012fast, wang2023resisting, wang2023meta, wang2023tdan}. As the collection of data, represented as graphs, reaches unprecedented levels, it becomes necessary to design efficient parallel algorithms for link prediction on such graphs. While link prediction algorithms are often pleasingly parallel, most studies do not address the design of suitable data structures for efficient computation of scores.

Further, the link prediction problem faces significant imbalance, with the number of known links often being several orders of magnitude less than non-existent links. This imbalance hinders the effectiveness of many link prediction methods, particularly on large networks \cite{wang2014link, garcia2014link}. Thus, heuristics are needed to minimize the computation needed, without sacrificing on quality.\ignore{To assess the accuracy of link prediction algorithms, observed links $E$ are split into a training set $E^T$ and a probe set $E^P$ for evaluation.} Quality assessment measures for link prediction include precision, recall, F1 score, accuracy, and Area Under the Receiver Curve (AUC). While AUC is commonly used \cite{arrar2023comprehensive}, it may provide misleading results\ignore{, by giving high scores to algorithms that successfully rank many negatives in the bottom} \cite{yang2015evaluating, lichtnwalter2012link}, leading to our focus on the F1 score.

\ignore{How do you explore the neighbors of each node, and compute intersection? What is a super naive way to do the above?}

\subsection{Our Contributions}

In this\ignore{technical} report, we study parallel algorithms for efficient link prediction in large graphs using neighborhood-based similarity measures. First, we improve upon the baseline approach (IBase). It efficiently finds common neighbors and handles large graphs by tracking top-$k$ edges per-thread and later merges them globally. Next, we propose a novel heuristic approach that additionally discards large hubs (DLH), based on the observation that high-degree nodes contribute poorly to similarity among their neighbors.\footnote{\url{https://github.com/puzzlef/neighborhood-link-prediction-openmp}} We then\ignore{experimentally} determine suitable hub limits, i.e, the degree above which a vertex is considered\ignore{as} a large hub, for link prediction with each similarity metric.

On a machine with two 16-core Intel Xeon Gold 6226R processors, our results show that the DLH approach outperforms IBase by over $1622\times$ and $415\times$ on average with $10^{-2}|E|$ and $0.1|E|$ unobserved edges, respectively. This speedup is achieved while maintaining comparable F1 scores. Notably, DLH achieves a link prediction rate of $38.1M$ edges/s with $0.1|E|$ unobserved edges. We also identified suitable similarity metric for each type of graph. When predicting $0.1|E|$ edges with the DLH approach, we observed that $63\%$ of the runtime is spent on the scoring phase, and especially higher on social networks, with high average degree. With doubling of threads, DLH exhibits an average performance scaling of $1.6\times$.

\section{Related work}
\label{sec:related}
Link prediction in network analysis encompasses various algorithms, such as similarity-based methods, dimensionality reduction, and machine learning \cite{arrar2023comprehensive}. The use of similarity measures for link prediction is based on the intuition that individuals tend to form connections with others who share similar characteristics \cite{wang2014link}. Similarity measures are broadly categorized into three types: local, quasi-local, and global. Local measures are based on the neighborhood information of nodes within a path distance of two, global measures consider information from the entire network for nodes, and quasi-local measures blend both approaches. Other similarity-based approaches involve random walks and communities \cite{arrar2023comprehensive}.

Dimensionality reduction techniques \cite{coskun2015link} attempt to map the network's information into a lower dimensional space, while preserving its structural information. These include embedding-based and matrix factorization based methods. On the other hand, machine learning techniques utilize extracted features from the network data to predict the probability of a link forming between two nodes based on these features. These include the use of supervised, unsupervised, reinforcement, and deep learning \cite{cui2018survey, arrar2023comprehensive}.

While effective in capturing nonlinear relationships for accurate predictions in complex networks, embedding-based methods face challenges with high-centrality nodes. Matrix factorization methods often struggle with complex network structures, require substantial computational resources and pose overfitting risks \cite{arrar2023comprehensive, martinez2016survey}. Machine learning based approaches have their own set of drawbacks. These include the need to obtain large labeled datasets, high-quality features, and domain expertise in supervised learning. Deep learning, while not requiring feature extraction, still demands a substantial amount of labeled data and raises concerns about interpretability and overfitting \cite{cui2018survey}.

Similarity-based link prediction methods continue to stay relevant, despite their usually lower prediction accuracy. This is because the size of graphs originating from the web, social networks or biological relations force us to use very simple algorithms if those graphs are to be computed in acceptable time \cite{garcia2014link}. Similarity-based link prediction methods are highly cost-effective, as they offer competitive prediction quality with their low complexity in time and space \cite{zhou2021progresses}. Further, in certain applications like friend recommendation, preference is given to predictions with explanations, a feature not readily achievable through machine learning techniques \cite{barbieri2014follow}.

Yang et al. \cite{yang2015new} introduce the Local Neighbors Link (LNL) measure, motivated by cohesion between common neighbors and predicted nodes, and implemented it on both MapReduce and Spark. Cui et al. \cite{cui2016bounded} present a parallel algorithm for efficiently evaluating Common Neighbors (CN) similarity, obtaining node pairs with CN values surpassing a specified lower bound. Guo et al. \cite{guo2019node} propose Common Neighbour Tightness (CNT), incorporating the aggregation degree of common neighbors by considering their proximity through local information and neighborhood tightness. Rafiee et al. \cite{rafiee2020cndp} introduce Common Neighbors Degree Penalization (CNDP), which factors in the clustering coefficient as a structural property and considers neighbors of shared neighbors. Mumin et al. \cite{mumin2022efficient} contribute an algorithm combining common neighbors and node degree distribution to estimate link presence likelihood between two nodes based on local information.

Papadimitriou et al. \cite{papadimitriou2012fast} introduce a similarity-based algorithm employing traversals on paths of limited length, grounded in the small world hypothesis. Their approach extends to directed and signed graphs, with discussions on a potential MapReduce implementation. Kalkan and Hambiralovic \cite{kalkanfinding} propose link prediction based on Personalized PageRank. Vega-Oliveros et al. \cite{vega2021link} investigate the use of susceptible-infected-recovered and independent cascade diffusion models. Their progressive-diffusion (PD) method, founded on nodes' propagation dynamics, provides a stochastic discrete-time rumor model for link prediction.

Mohan et al. \cite{mohan2017scalable} introduce a hybrid similarity measure utilizing parallel label propagation for community detection and a parallel community information-based Adamic-Adar measure, employing the Bulk Synchronous Parallel (BSP) programming model. Wang et al. \cite{wang2019link} propose a link prediction algorithm incorporating an adjustable parameter based on community information (CI), applying it to various similarity indices and a family of CI-based indices. They also develop a parallel algorithm for large-scale complex networks using Spark GraphX. Bastami et al. \cite{bastami2019gravitation} present a gravitation-based link prediction approach, enhancing local and global predictions through the integration of node features, community information, and graph properties. Saifi et al. \cite{saifi2023fast} propose an approach that accelerates link prediction using local and path-based similarity measures by operating on the connected components of a network rather than the entire network.

Shin et al. \cite{shin2012multi} introduce Multi-Scale Link Prediction (MSLP), employing a tree-structured approximation algorithm for efficient link prediction in large networks. Garcia-Gasulla and Cort{\'e}s \cite{garcia2014link} propose a local link prediction algorithm based on an underlying hierarchical model, emphasizing aspects of parallelization, approximation, and data locality for computational efficiency. Ferreira et al. \cite{ferreira2019scalability} present a multilevel optimization to enhance the scalability of any link prediction algorithm by reducing the original network to a coarsened version. Benhidour et al. \cite{benhidour2022approach} propose a link prediction method for directed networks, leveraging the similarity-popularity paradigm. The algorithms approximate hidden similarities as shortest path distances, using edge weights that capture and factor out links' asymmetry and nodes' popularity.

As mentioned earlier, similarity-based algorithms are competitive with other high-quality dimensionality reduction and machine learning techniques, thanks to their simplicity, interpretability \cite{pai2019netdx, barbieri2014follow}, and computational efficiency \cite{garcia2014link} --- and are thus often combined with other techniques \cite{kumari2022supervised, abuoda2020link, pai2019netdx}. However, the evaluation of these algorithms on large networks is crucial, as testing on small networks can yield misleading conclusions \cite{zhou2021progresses, zhou2021experimental}. Despite this, a significant portion of the discussed works focuses on small \cite{guo2019node, rafiee2020cndp, mumin2022efficient, papadimitriou2012fast, vega2021link, saifi2023fast, ferreira2019scalability, benhidour2022approach, wang2023resisting, wang2023meta, wang2023tdan} to medium-scale graphs \cite{yang2015new, cui2016bounded, kalkanfinding, mohan2017scalable, wang2019link, bastami2019gravitation, shin2012multi, garcia2014link}, with less than a million or billion edges. Parallelism becomes essential on large networks, and while some approaches, such as the ones based on common neighbors \cite{yang2015new, cui2016bounded}, random walks \cite{papadimitriou2012fast}, community structures \cite{mohan2017scalable, wang2019link}, and approximation \cite{garcia2014link}, incorporate parallelism, the design of suitable data structures for efficient score computation remains an often-overlooked aspect. This technical report aims to bridge both gaps, while proposing a heuristic for efficient computation.

\section{Preliminaries}
\label{sec:preliminaries}
In an undirected graph $G(V, E)$ with sets of vertices $V$ and edges $E$, link prediction aims to identify missing or future edges from the set $U - E$, where $U$ contains all possible $|V|(|V|-1)/2$ edges in the graph \cite{zhou2021progresses}. Link prediction, thus, is akin to finding a needle in a haystack, as the correct edges need to be identified within a vast set of incorrect ones \cite{garcia2014link, wang2014link}. Garcia et. al \cite{garcia2014link} observe that this ratio goes from $1:11k$ in their best case to $1:27M$ in the worst case. Further, larger graphs are more likely to be incomplete. These challenges make the study of link prediction crucial.

Link prediction often relies on similarity metrics between node pairs, reflecting the likelihood of missing or future links \cite{wang2014link, arrar2023comprehensive}. The idea is grounded in the tendency for users to connect with similar individuals. More similarity thus suggests a higher probability of a future link. A ranked list of potential links, based on similarity scores, can then be used to predict the top-$k$ links that are most likely to appear (or were missing) \cite{wang2014link}. Similarity measures are commonly categorized into local, quasi-local, and global measures. Local / neighborhood-based metrics, like Common Neighbours (CN), are calculated based on neighborhood information within a path distance of two. Global indices use network-wide information, while quasi-local indices combine both for distances up to two \cite{arrar2023comprehensive}.

As brought up earlier, despite typically having lower prediction accuracy than machine learning based approaches, similarity-based link prediction methods remain relevant due to the need for simple algorithms in handling large graphs \cite{garcia2014link}. Further, they are highly cost-effective, interpretable, and offer competitive prediction quality with low time and space complexity \cite{zhou2021progresses, barbieri2014follow}.

\subsection{Neighborhood-based Similarity Metrics}
\label{sec:metrics}

We now discuss nine commonly used local / neighborhood-based similarity metrics for link prediction.

\subsubsection{Common Neighbors (CN)}

This metric, shown in Equation \ref{eq:cn}, counts the shared neighbors between two vertices, $a$ and $c$, in a graph \cite{newman2001clustering}. However, its lack of normalization may pose challenges when comparing node pairs with different degrees of connectivity.

\begin{equation}
\label{eq:cn}
  CN(a, c) = |\Gamma_a \cap \Gamma_c|
\end{equation}

\subsubsection{Jaccard Coefficient (JC)}

The Jaccard Coefficient (JC) \cite{jaccard1901etude} is a popular similarity measure\ignore{in network analysis and link prediction}. It offers a normalized assessment of similarity between nodes based on their neighborhoods. Defined by Equation \ref{eq:jc}, JC assigns higher values to node pairs with a greater proportion of common neighbors relative to their total neighbors.

\begin{equation}
\label{eq:jc}
  JC(a, c) = \frac{|\Gamma_a \cap \Gamma_c|}{|\Gamma_a \cup \Gamma_c|}
\end{equation}

\subsubsection{S{\o}rensen Index (SI)}

S{\o}rensen Index (SI) \cite{sorensen1948method}, also known as S{\o}rensen–Dice coefficient, is another similarity metric commonly applied in network analysis and link prediction. This metric, defined by Equation \ref{eq:si}, extends beyond solely accounting for the size of common neighbors and introduces the idea that nodes with lower degrees are more likely to form links.

\begin{equation}
\label{eq:si}
  SI(a, c) = \frac{|\Gamma_a \cap \Gamma_c|}{|\Gamma_a| + |\Gamma_c|}
\end{equation}

\subsubsection{Salton Cosine similarity (SC)}

The Salton Cosine similarity (SC) \cite{salton1973specification} essentially measures the cosine of the angle between the vectors representing the neighborhoods of nodes $a$ and $c$, as given in Equation \ref{eq:sc}. Similar to other metrics, a higher SC value indicates a greater similarity in the neighborhood structures of the nodes, implying a higher likelihood of a link between them.

\begin{equation}
\label{eq:sc}
  SC(a, c) = \frac{|\Gamma_a \cap \Gamma_c|}{\sqrt{|\Gamma_a| \cdot |\Gamma_c|}}
\end{equation}

\subsubsection{Hub Promoted (HP)}

The HP \cite{liben2003link} metric, defined by Equation \ref{eq:hp}, assesses topological overlap between two nodes in a graph. It is particularly influenced by the lower degree of nodes, and can be valuable in scenarios where the involvement of lower-degree nodes is considered important in understanding network connectivity.

\begin{equation}
\label{eq:hp}
  HP(a, c) = \frac{|\Gamma_a \cap \Gamma_c|}{min(|\Gamma_a|, |\Gamma_c|)}
\end{equation}

\subsubsection{Hub Depressed (HD)}

In contrast to the HP metric, the Hub Depressed (HD) score \cite{zhou2009predicting} is determined by the higher degrees of nodes, as illustrated by Equation \ref{eq:hd}. It can be particularly useful in cases where the influence of highly connected nodes on network structure is of interest.

\begin{equation}
\label{eq:hd}
  HD(a, c) = \frac{|\Gamma_a \cap \Gamma_c|}{max(|\Gamma_a|, |\Gamma_c|)}
\end{equation}

\subsubsection{Leicht-Holme-Nerman (LHN)}

The Leicht-Holme-Nerman (LHN) score \cite{leicht2006vertex} is a similarity metric that assigns high similarity to node pairs that exhibit a greater number of common neighbors than would be expected by random chance. One may use Equation \ref{eq:lhn} to compute the LHN score between two nodes $a$ and $b$ in a graph.

\begin{equation}
\label{eq:lhn}
  LHN(a, c) = \frac{|\Gamma_a \cap \Gamma_c|}{|\Gamma_a| \cdot |\Gamma_c|}
\end{equation}

\subsubsection{Adamic-Adar coefficient (AA)}

AA \cite{adamic2003friends} is a popular measure designed to capture the notion that connections to common neighbors with fewer links are more informative and indicative of similarity between nodes in a network. The formula in Equation \ref{eq:aa} assigns weights inversely proportional to the logarithm of the number of neighbors, reducing sensitivity to highly connected nodes.\ignore{AA highlights that links to less common neighbors provide more discriminative information about node similarity.}

\begin{equation}
\label{eq:aa}
  AA(a, c) = \sum_{b\ \in\ \Gamma_a \cup \Gamma_c} \frac{1}{\log{|\Gamma_b|}}
\end{equation}

\subsubsection{Resource Allocation (RA)}

The RA metric \cite{zhou2010solving} is based on the concept of heat diffusion in a network, emphasizing that heavily connected nodes may not play a critical role in facilitating resource flow between other nodes. Unlike AA, RA penalizes high-degree common neighbors more heavily. The score between nodes $a$ and $c$ is determined by Equation \ref{eq:ra}.

\begin{equation}
\label{eq:ra}
  RA(a, c) = \sum_{b\ \in\ \Gamma_a \cup \Gamma_c} \frac{1}{|\Gamma_b|}
\end{equation}

Note that the CN, AA, and RA metrics lack normalization, and thus only convey ranking information \cite{wang2014link}. In practical applications, one should choose the right metric based on the network's characteristics --- there is no universally dominating metric \cite{zhou2021progresses, ghasemian2020stacking, wang2014link, liben2003link}.

\subsection{Measuring Prediction Quality}

A number of measures are used to assess link prediction performance. These include precision, recall, F1 score, and Area Under the Receiver Curve (AUC) \cite{arrar2023comprehensive}. However, AUC is insufficient for early quality assessment, as it may inaccurately favor algorithms ranking many negatives at the bottom \cite{zhou2021progresses, yang2015evaluating, lichtnwalter2012link}. Accordingly, we focus on precision, recall, and F1 score in this report \cite{lu2015toward}.

\subsubsection{Precision}

Precision measures the ratio of correctly predicted links $P^\checkmark$ to all predicted links $P = P^\checkmark \cup P^\times$, where $P^\times$ is the set of incorrectly predicted links \cite{arrar2023comprehensive, zhou2021progresses}. Equation \ref{eq:precision} provides the formula for precision computation.

\begin{equation}
\label{eq:precision}
  \text{Precision} = \frac{|P^\checkmark|}{|P|}
\end{equation}

\subsubsection{Recall}

Unlike precision, recall measures the ratio of correctly predicted links $P^\checkmark$ to the new ground-truth links $E^U$, as presented in Equation \ref{eq:recall} \cite{arrar2023comprehensive, zhou2021progresses}. If the number of predicted links is equal to the number of new ground-truth links, then recall is the same as precision \cite{zhou2021progresses, lu2011link, liben2003link}.

\begin{equation}
\label{eq:recall}
  \text{Recall} = \frac{|P^\checkmark|}{|E^U|}
\end{equation}

\subsubsection{F1 score}

F1 score is a measure of the balance between precision and recall \cite{arrar2023comprehensive}. It is calculated as the harmonic mean of precision and recall, as shown in Equation \ref{eq:f1score}.

\begin{equation}
\label{eq:f1score}
  \text{F1 score} = 2 * \frac{\text{Precision} * \text{Recall}}{\text{Precision} + \text{Recall}}
\end{equation}

A majority of link prediction studies apply random division of the ground-truth edges $E$, into sets of observed $E^O$ and unobserved edges $E^U$ to assess link prediction performance \cite{zhou2021progresses}. The number of unobserved edges $|E^U|$ is commonly set at $10\%$ of total links in $E$, based on empirical findings that it yields statistically solid results without significantly altering the network's structure \cite{lu2015toward}.

\subsection{Baseline approach for Neighborhood-based Link Prediction}

A baseline approach for neighborhood-based link prediction involves computing similarity scores between all non-connected node pairs $\{(a, c)\ |\ a, c \in V; (a, c) \notin E\}$, and returning the node pairs with top-$k$ scores as the predicted links. The similarity score is computed by assessing the commonality of neighbors between the two nodes. This approach is used by a number of studies previously mentioned \cite{gatadi2023lpcd, saifi2023fast, benhidour2022approach, mumin2022efficient, rafiee2020cndp, guo2019node, yang2015new, papadimitriou2012fast, wang2019link}. Notably, popular network analysis software packages such as NetworKit \cite{staudt2016networkit} and igraph \cite{csardi2006igraph} also implement this baseline approach.

While this approach is simple to understand, it has severe computational costs. Finding the common neighbors $\Gamma_a \cap \Gamma_c$ of node pairs $(a, c)$ using a naive method has a time complexity of $O(D^2)$, where $D$ is the degree of the maximum degree node in the graph. This results in an overall time complexity of $O(N^2D^2)$. Using a hashtable, one can only reduce the time complexity to $O(N^2D)$. Additionally, this method involves significant unnecessary computations, as many node pairs are likely to lack common neighbors.

\section{Approach}
\label{sec:approach}
\subsection{Our Improved Baseline approach (IBase)}

\subsubsection{Avoiding node pairs with no common neighbors}

As noted earlier, there are a vast number of node pairs $(a, c)$ in a graph with no common neighbors, i.e., $\Gamma_a \cap \Gamma_c = \phi$ (here $\Gamma_a$ represents the neighbors of vertex $a$), and thus have a similarity score of $0$. This is especially true for large, sparse graphs. To address this, we limit the computation of similarity scores of each vertex $a$ in the graph to its second-order neighbors $c$, i.e., to vertices that are neighbors of the immediate neighbors $a$. We do this with a Depth First Search (DFS) traversal from each vertex $a$, limited to a distance of $2$.

\subsubsection{Finding common neighbors faster}

To expedite the identification of common neighbors between each vertex $a$ and its second-order neighbors $c$, we employ a hashtable. This hashtable keeps track of the frequency of visits to each second-order vertex $c$ during the DFS traversal, yielding the count of common neighbors $|\Gamma_a \cap \Gamma_c|$ in each hashtable entry (with $c$ as the key). This count reflects the number of paths from $a$ to $c$, as each common neighbor contributes a new path. For the JC metric, in addition to common neighbors, we need the total count of neighbors $\Gamma_a \cup \Gamma_c$ between $a$ and $c$. This count is easily calculated as $|\Gamma_a \cup \Gamma_c| = |\Gamma_a| + |\Gamma_c| - |\Gamma_a \cap \Gamma_c|$. To compute AA and RA similarity scores, we use a hashtable with floating-point values, and accumulate $1/\log{|\Gamma_b|}$ and $1/|\Gamma_b|$ respectively. Here, $b$ is a first-order neighbor of $a$ and a common neighbor between $a$ and $b$. Finally, to avoid redundancy in an undirected graph, we skip second-order neighbors $c$ where $c \leq a$. This also prevents the scoring (and prediction) of self-loops.

\subsubsection{Design of Hashtable}

\begin{figure}[hbtp]
  \centering
  \subfigure{
    \label{fig:about-hashtable--all}
    \includegraphics[width=0.88\linewidth]{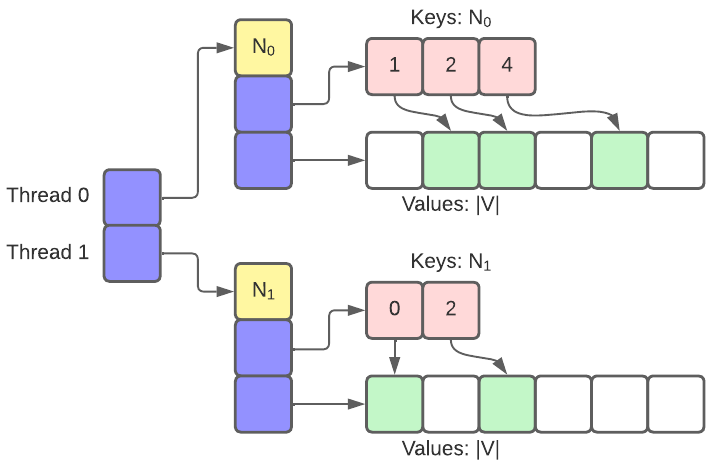}
  } \\[-2ex]
  \caption{Illustration of collision-free per-thread hashtables that are well separated in their memory addresses, for two threads. Each hashtable consists of a keys vector, a values vector (of size $|V|$), and a key count ($N_0$/$N_1$). The value associated with each key is stored/accumulated in the index pointed by the key. As the key count of each hashtable is updated independently, we allocate it separately on the heap to avoid false cache sharing \cite{sahu2023gvelouvain}. These are used in the scoring phase of our implementation to track common neighbors.}
  \label{fig:about-hashtable}
\end{figure}

As C++'s inbuilt \texttt{map} has poor performance, we employ collision-free per-thread hashtables (we discuss parallelization later), similar to our previous work \cite{sahu2023gvelouvain}. An example of this hashtable for two threads is shown in Figure \ref{fig:about-hashtable}. Each hashtable includes a keys vector, a values vector (of size $|V|$), and a key count. The value associated with each key is stored or accumulated in the index pointed to by the key. To prevent false cache sharing, we independently update the key count of each hashtable and allocate it separately on the heap. As previously mentioned, the vertex IDs of second-order neighbors $c$ of each vertex $a$ serve as keys in the hashtable, and the associated values represent the number of times $c$ has been visited during a DFS traversal (of distance $2$) from $a$. For AA and RA scores, we use a hashtable with integral keys and floating-point values.

\subsubsection{Avoiding first-order neighbors}

Once the DFS traversal from vertex $a$ is complete, and all hashtable entries are populated, we clear the entries associated with the first-order neighbors $b$ of $a$. This precaution is necessary because some first-order neighbors of $a$ may have edges with the other first-order neighbors.

\subsubsection{Computing scores}

With all valid entries in the hashtable populated, we proceed to compute scores from vertex $a$ to each second-order neighbor $c$, using the appropriate formula for the selected metric as outlined in Section \ref{sec:metrics}. For CN, AA, and RA metrics, no additional computation is required, as the hashtable values already contain the desired scores.

\subsubsection{Tracking top-k edges}

Since storing all obtained similarity scores is impractical due to the scale of processed graphs, we use a min-heap based prediction list. This min-heap enables us to maintain the top-$k$ edges with the highest similarity scores per thread. It works by evicting the node pair with the lowest score when a new node pair with a higher score is encountered. As an optimization, we convert the per-thread prediction lists into a min-heap only after they have been populated with $k$ entries.

\subsubsection{Parallelizing the process}

To parallelize the computation, we partition the graph among threads, employing OpenMP's \textit{dynamic} schedule with a chunk size of $2048$ (to minimize work imbalance and scheduling overhead). Each partition is processed independently, with each thread using its own hashtable and prediction list. To account for the possibility of a single thread identifying all top-$k$ links globally, each thread employs a prediction list of size $k$. After individual thread predictions, we independently sort the per-thread top-$k$ predictions by score on each thread. These sorted lists are then merged into a global top-$k$ prediction list using a max-heap.

\ignore{An alternative approach to merge per-thread top-$k$ prediction lists involves concatenating the per-thread lists into a single prediction list, performing a parallel sort, and truncating all predicted edges except the top-$k$. However, this method can lead to out-of-memory crashes on large graphs due to the size of the concatenated prediction list.}

We call this approach IBase. It has a time complexity of $O(ND^2)$, where $N$ is the number of vertices in the graph, and $D$ is the maximum degree of a vertex in the graph.\ignore{It's worth noting that the results obtained by our parallel algorithm are non-deterministic. This is because multiple edges can have matching scores, and the order of edges with the same score may be random due to parallelism.}

\subsection{Our Disregard Large Hubs (DLH) heuristic}

\subsubsection{Explanation of the approach}

To optimize link prediction further, we recognize that low-degree nodes, representing users with few connections in the social network, are more selective in accepting friend requests and are likely to form connections with people they have stronger, more meaningful relationships with, such as close friends and family. Thus, low-degree nodes confer significant similarity among their neighbors, while high-degree nodes generally do not (due to their lack of selectivity). We refer to such high-degree nodes as large hubs. Accordingly, for a given vertex $a$ and each of its immediate neighbors $b$, we only explore the neighbors $c \in \Gamma_b$ of $b$ if the degree of $b$ is within a certain threshold $L_H$, i.e., $|\Gamma_b| \leq L_H$. We call this threshold the \textit{hub limit}. This minimizes DFS exploration of second-degree neighbors $c$ during the computation of neighbor-based similarity scores.

We call this approach DLH. It has the same time complexity as IBase, i.e., $O(ND^2)$, but significantly outperforms it by runtime, while achieving the same or better prediction quality.

\subsubsection{A simple example}

\begin{figure*}[hbtp]
  \centering
  \subfigure[Our Improved Baseline approach (IBase)]{
    \label{fig:about-pruning--01}
    \includegraphics[width=0.31\linewidth]{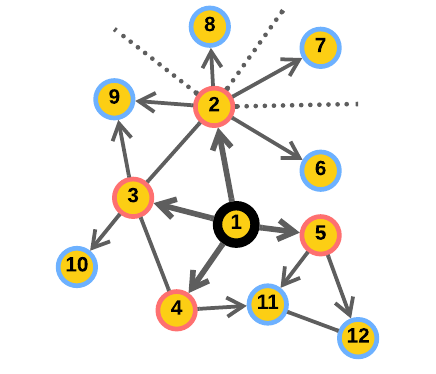}
  }
  \subfigure[Disregard hubs with degree $> 8$ (DLH)]{
    \label{fig:about-pruning--02}
    \includegraphics[width=0.31\linewidth]{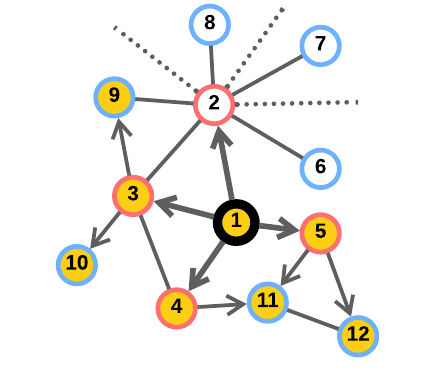}
  }
  \subfigure[Disregard hubs with degree $> 4$ (DLH)]{
    \label{fig:about-pruning--03}
    \includegraphics[width=0.31\linewidth]{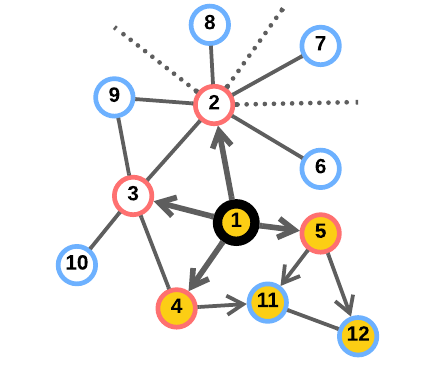}
  } \\[-2ex]
  \caption{Illustration of our\ignore{neighborhood-based} link prediction approach which \textit{Disregards Large Hubs (DLH)}, i.e., $1^{st}$ order neighbors with high degree\ignore{(DLH)}. Here we focus on the neighborhood of vertex $1$\ignore{in the graph}, but the approach applies to each vertex in the graph. In the figure,\ignore{the current vertex} $1$ is outlined in black, its $1^{st}$ order neighbors in red, its $2^{nd}$ order neighbors in blue, and explored/traversed vertices are shown with a yellow fill. Edge directions indicate traversal, with some $2^{nd}$ order vertices omitted for simplicity (dotted edges). (a) Depicts our \textit{Improved Baseline (IBase)} approach, which considers all $2^{nd}$ order neighbors of vertex $1$. (b) Presents our DLH approach, which considers only $2^{nd}$ order neighbors linked to $1$ through a small hub (degree $\leq 8$). This pruning reduces runtime and enhances prediction quality. (c) Illustrates the DLH approach, where vertices with degree $> 4$ are considered large hubs.}
  \label{fig:about-pruning}
\end{figure*}

Figure \ref{fig:about-pruning} demonstrates the DLH approach. Here, we consider the neighborhood of vertex $1$ in a graph, where $1$ is outlined in black, its first-order neighbors in red, its second-order neighbors in blue, and explored/traversed vertices are shown with a yellow fill. The arrows in the figure indicate the direction of DFS traversal, and for figure simplicity, three neighbors of vertex $2$ are omitted (shown with dotted edges).

\ignore{\paragraph{Standard approach (IBase)}}

Figure \ref{fig:about-pruning--01} depicts the standard approach, which considers all second-order neighbors of vertex $1$ for score computation, i.e., vertices $6$ to $12$. This is the process followed by the IBase approach.

\ignore{\paragraph{Disregards hubs with degree $> 8$ (DLH)}}

Figure \ref{fig:about-pruning--02} depicts our DLH approach. It considers only second-order neighbors linked to vertex $1$ by a small hub, with a degree $\leq 8$, for score computation --- i.e., vertices $9$ to $12$, which are linked to $1$ through vertices $3$, $4$, and $5$. It is based on the idea, as mentioned above, that low-degree vertices contribute more similarity among their neighbors, while vertices with high degrees do not.\ignore{This minimization of exploration significantly reduces runtime and enhances prediction quality, as discussed in Section \ref{sec:select-limit}.} Note that second-order neighbors of $1$ can still be considered if linked to $1$ by a small hub (e.g., vertex $9$ linked to $1$ through $3$).

\ignore{\paragraph{Disregards hubs with degree $> 4$ (DLH)}}

Lastly, Figure \ref{fig:about-pruning--03} shows the DLH approach, considering only second-order neighbors linked to vertex $1$ by a small hub with a degree $\leq 4$ ($11$ and $12$ linked to $1$ through $4$ and $5$). First-order neighbors of $1$ with a degree $> 4$ are considered large hubs, and thus their second-order neighbors are disregarded.\ignore{This approach is applied to all vertices in the graph. Note that Figure \ref{fig:about-pruning} only shows the neighborhood of vertex $1$.}

\subsubsection{Non-determinism in the result}

It is worth noting that the results obtained by our parallel algorithms (IBase and DLH) are non-deterministic. This is because multiple edges can have matching scores, and the order of edges with the same score may be random due to parallelism.

\subsection{Framework of our IHub and DLH approaches}

\begin{figure*}[hbtp]
  \centering
  \subfigure{
    \label{fig:about-framework--all}
    \includegraphics[width=0.98\linewidth]{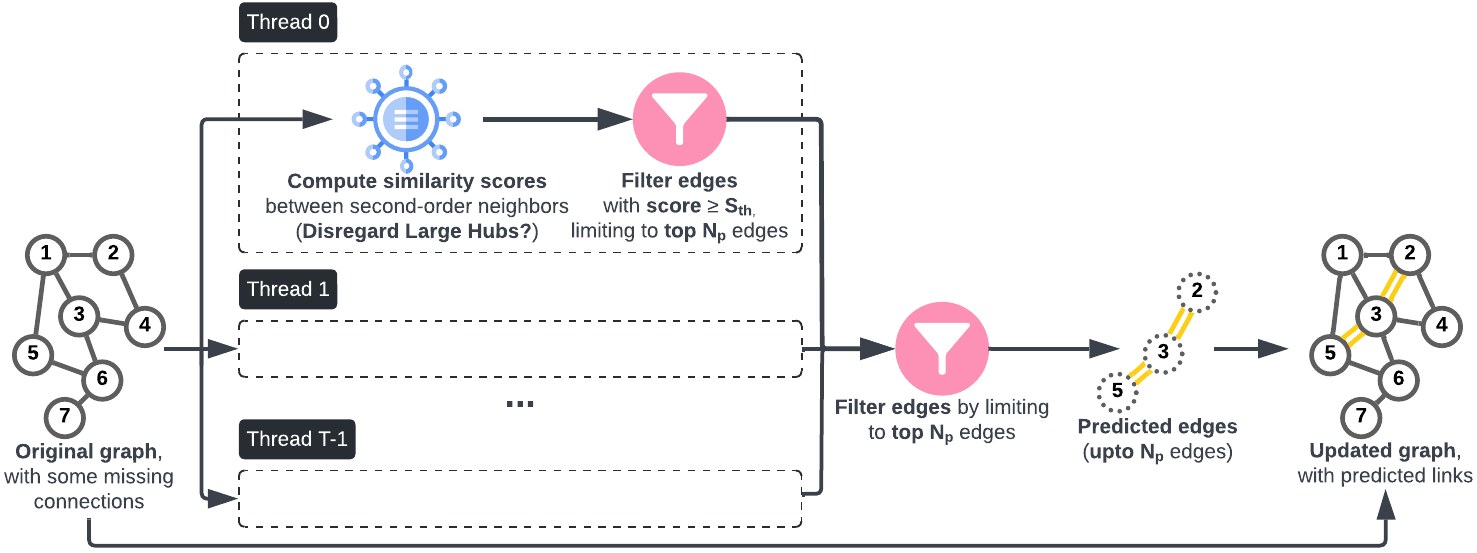}
  } \\[0ex]
  \caption{Framework diagram of our \textit{IBase} and \textit{DLH} approaches for link prediction. Here, $N_p$ denotes the maximum number of edges to predict, while $S_{th}$ indicates the threshold score for making a prediction.}
  \label{fig:about-framework}
\end{figure*}

Figure \ref{fig:about-framework} shows the framework of our IBase and DLH approaches. Given an original graph with missing edges, $T$ threads partition the vertices into $T$ sections (using dynamic scheduling). Each thread computes similarity scores between non-connected second-order neighbors for each source vertex, with the option to ignore large hubs (in the DLH approach). Each thread filters the computed scores, retaining only the top $N_p$ edges with scores $\geq S_{th}$. The filtered edges are then merged and globally filtered again to keep the top $N_p$ edges overall, yielding the predicted edges. These predicted edges can be added to the original graph to obtain an updated graph, which can then be used for further analysis.

\subsection{Selecting Suitable Hub Limit for each Similarity metric}
\label{sec:select-limit}

We now select a suitable hub limit $L_H$, i.e., a degree above which a vertex is considered a large hub, for each similarity metric presented in Section \ref{sec:metrics}. For this, we adjust the hub limit $L_H$ from $2$ to $1024$ (in multiples of $2$), for each similarity metric, and adjust the number of unobserved edges $E^U$ from $10^{-4}|E|$ to $0.1|E|$ on a number of graphs. We then apply each similarity metric based link prediction to predict $N_P = |E^U|$ edges with the highest similarity scores. We also test with a hub limit $L_H$ of $\infty$, which is\ignore{essentially} the IBase approach.

\begin{figure*}[hbtp]
  \centering
  \subfigure[Relative runtime (logarithmic scale) of each link prediction method]{
    \label{fig:adjust-mindegree--runtime}
    \includegraphics[width=0.98\linewidth]{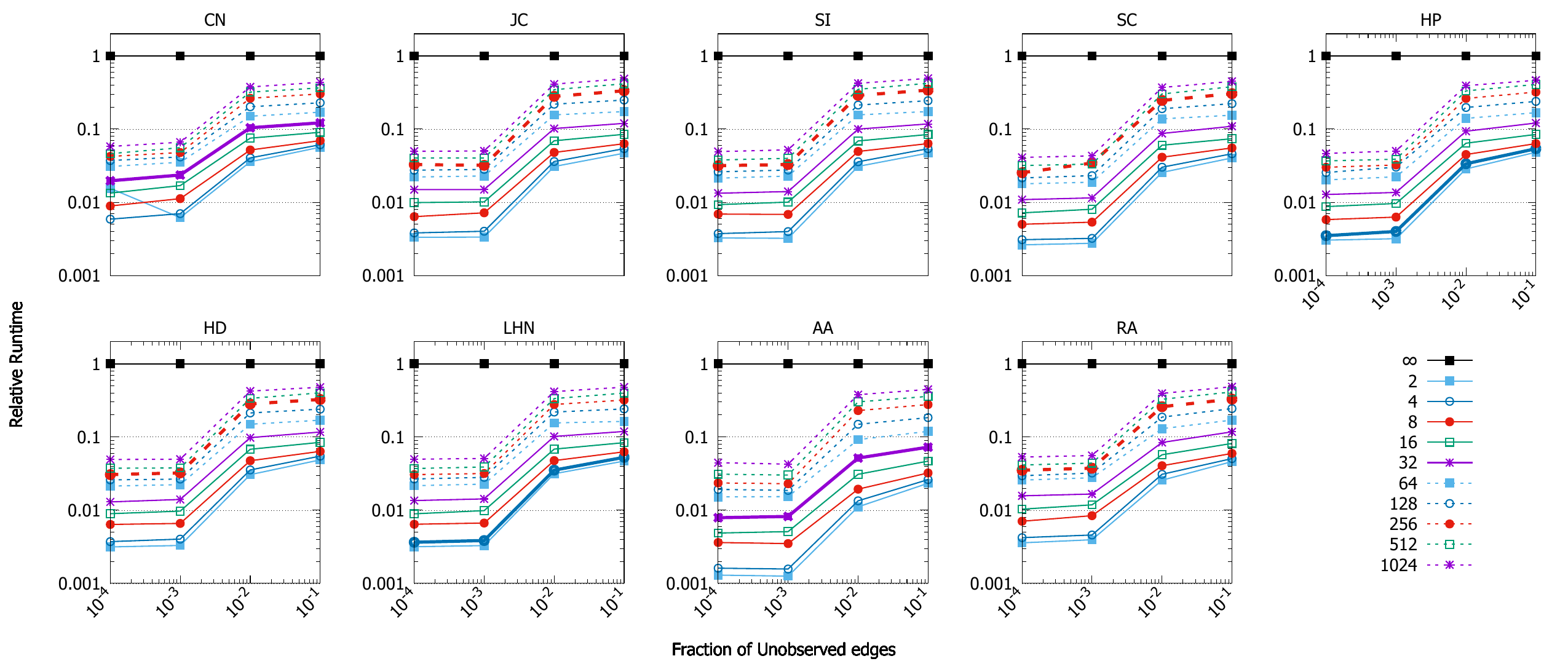}
  }
  \subfigure[F1 score of predicted links (logarithmic scale) of each link prediction method]{
    \label{fig:adjust-mindegree--precision}
    \includegraphics[width=0.98\linewidth]{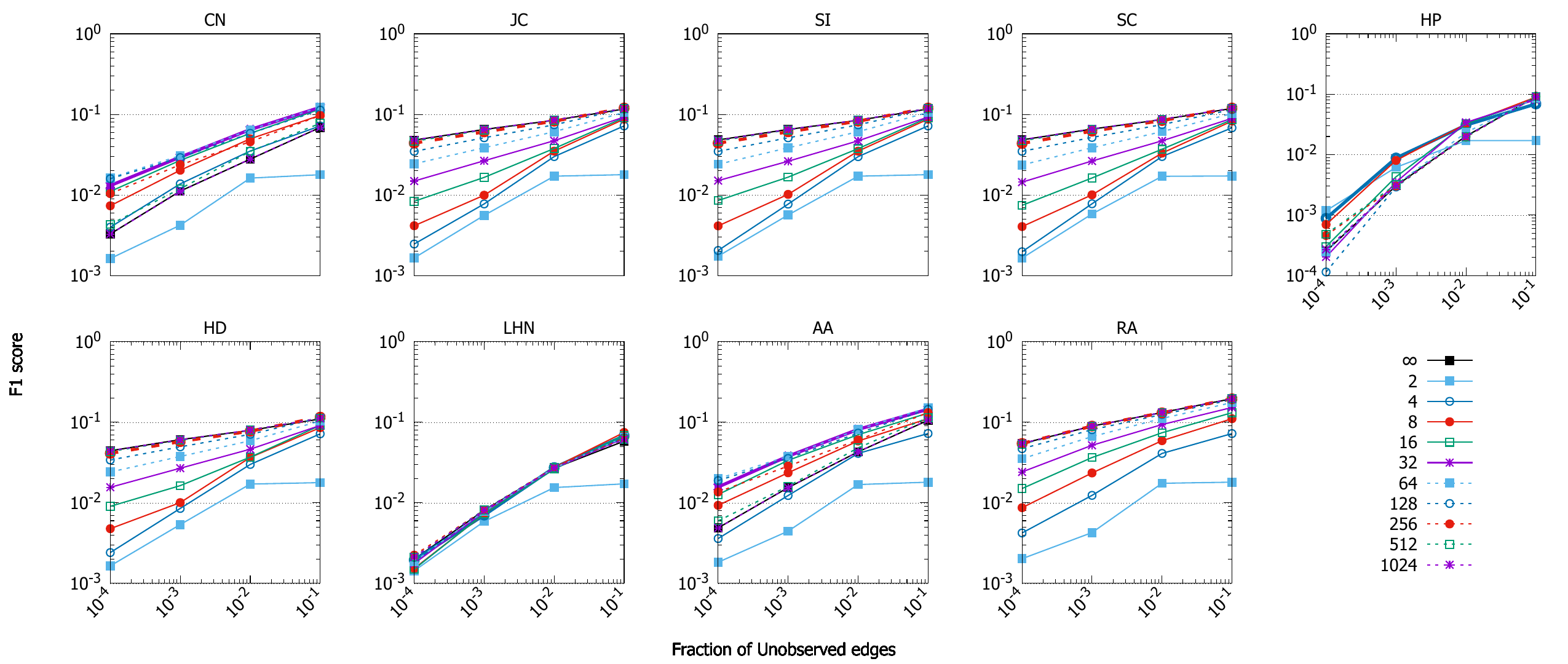}
  } \\[0ex]
  \caption{Impact of adjusting the hub limit $L_H$ from $2$ to $1024$ (in multiples of $2$), and to $\infty$ (\textit{IBase} approach), on the runtime (log-scale), and F1 score of predicted links (log scale), of each neighbor-based link prediction method, with the number of unobserved edges $E^U$ ranging from $10^{-4}|E|$ to $0.1|E|$. The full form of each link prediction method is given in Section \ref{sec:metrics}.}
  \label{fig:adjust-mindegree}
\end{figure*}

Figure \ref{fig:adjust-mindegree--runtime} shows the mean runtime taken for each similarity metric, with each hub limit $L_H$, and with the number of unobserved edges $E^U$ ranging from $10^{-4}|E|$ to $0.1|E|$, while Figure \ref{fig:adjust-mindegree--precision} shows the mean F1 score of the predicted edges, with the IBase and DLH approaches. Results indicate that selecting a lower hub limit $L_H$ decreases the overall runtime of the DLH approach. In terms of both F1 score and relative runtime, we observe that a hub limit $L_H$ of $4$ is suitable for HP and LHN metrics, a hub limit $L_H$ of $32$ is suitable for CN and AA metrics, and a hub limit $L_H$ of $256$ is suitable for the remaining metrics (i.e., JC, SI, SC, HD, and RA). These hub limits are highlighted with thick lines in the figures. Further, as Figure \ref{fig:adjust-overall} shows, hub limits $L_H$ of $4$, $32$, and $256$ offer a mean speedup of $67\times$, $32\times$, and $13\times$ when the number of unobserved edges $E^U$ is $0.1|E|$. Indeed, disregarding large hubs, with the DLH approach, offers a significant improvement in performance with little to no degradation in the accuracy of link prediction.

\subsection{Our implementation of the Disregard Large Hubs (DLH) approach}

We now explain the implementation of our DLH approach for parallel neighborhood-based link prediction, which disregards large hubs. The pseudocode for this is given in Algorithm \ref{alg:predict}. Here, the \texttt{predictLinks} function accepts the input graph $G$, the maximum number of edges to predict $N_P$, a threshold score $S_{th}$, and outputs the list of predicted edges $P$. The algorithm operates in two phases: the scoring phase and the merging phase.

In the \textit{scoring phase} (lines \ref{alg:predict--scoring-begin}-\ref{alg:predict--scoring-end}), each vertex is processed in parallel. For each vertex $a$, we scan its second-order neighbors $c$, considering only neighbors of its first-order neighbors $b$, which have a degree $|\Gamma_b|$ less than or equal to the hub limit $L_H$\ignore{ --- this is based on the idea that large hubs (here, the first order neighbors) confer little similarity among its neighbors (here, between the current vertex $i$, and its second order neighbors)}. The scoring of potential edges is done in the \texttt{scanEdges} function (lines \ref{alg:predict--scan-begin}-\ref{alg:predict--scan-end}). Here, we skip the reverse edges (where $c \leq a$), and calculate a score for each potential second-order neighbor $c$ based on the given metric, i.e., for the AA and RA metrics, we apply the scoring formula for each common neighbor $b$, and for the other metrics, we simply count the number of common neighbors. The scores\ignore{, or the number of common neighbors} are accumulated in a collision-free per-thread hashtable $H_t$.

After scanning, we set entries corresponding to first-order neighbors of $a$ to $0$ in $H_t$ to avoid considering them as potential edges (line \ref{alg:predict--avoid-neighbors1}). We then calculate a score for each potential edge $(a, c)$ from the hashtable $H_t$, and if the score exceeds the specified threshold $S_{th}$, the edge is added to a per-thread prediction list $P_t$. The list is maintained as a min-heap based on scores (after $N_P$ edges have been added), in order to efficiently keep track of top $N_P$ edges. The size of $P_t$ is controlled to retain only the top-scoring edges. After scoring all the vertices in parallel, the merging phase begins.

In the \textit{merging phase} (lines \ref{alg:predict--merging-begin}-\ref{alg:predict--merging-end}), the per-thread prediction lists are merged into a global prediction list $P$. This is done by first sorting the per-thread lists based on the scores. A max heap $P_h$ is then created to track the maximum score from each thread. We then iteratively select the highest-scoring edge from the per-thread prediction lists, and add it to the global list until either the maximum number of edges is reached $N_P$ or there are no more edges to consider (lines \ref{alg:predict--merge-loop-begin}-\ref{alg:predict--merging-end}).

\begin{algorithm}[hbtp]
\caption{Our Parallel \textit{Disregard Large Hubs (DLH)} approach.}
\label{alg:predict}
\begin{algorithmic}[1]
\Require{$G(V, E)$: Input graph}
\Require{$N_P$: Maximum number of edges to predict}
\Require{$S_{th}$: Threshold score above which to predict}
\Ensure{$a, b, c$: Current vertex, first-order, second-order neighbor}
\Ensure{$L_H$: Hub limit, i.e., large hub degree threshold}
\Ensure{$|\Gamma_b|$: Degree of first-order neighbor $b$}
\Ensure{$H_t$: Collision-free per-thread hashtable}
\Ensure{$P_t$: Per-thread prediction list}
\Ensure{$P_h$: Heap for merging per-thread prediction lists}
\Ensure{$P$: Global prediction list}

\Statex

\Function{predictLinksDLH}{$G, N_P, S_{th}$} \label{alg:predict--begin}
  \State $\rhd$ Scoring phase
  \State $P_t \gets \{\}$ \textbf{on each thread} \label{alg:predict--scoring-begin}
  \ForAll{$a \in V$ \textbf{in parallel}}
    \State $\rhd$ Scan second-order neighbors of $a$
    \State $H_t \gets \{\}$
    \ForAll{$b \in G.out(a)$}
      \State $\rhd$ Skip high degree first-order neighbors
      \If{$|\Gamma_b| \leq L_H$} $scanEdges(H_t, G, b)$
      \EndIf
    \EndFor
    \State $\rhd$ Avoid first-order neighbors
    \ForAll{$b \in G.out(a)$} $H_t[b] \gets 0$ \label{alg:predict--avoid-neighbors1}
    \EndFor
    \State $\rhd$ Get prediction scores and add to list
    \ForAll{$c \in H_t.keys()$}
      \State $score \gets computeScore(a, c, H_t[c])$
      \If{$score \leq S_{th}$} \textbf{continue}
      \EndIf
      \State $\rhd$ Add edge $(a, c)$ to prediction list
      \If{$|P_t| < N_P$}
        \State $P_t.push(\{a, c, score\})$
        \If{$|P_t| = N_P$} $P_t.makeMinHeapByScore()$
        \EndIf
      \ElsIf{$score \geq P_t[0].score$}
        \State $P_t.popHeap()$
        \State $P_t.pushHeap(\{a, c, score\})$
      \EndIf
    \EndFor
  \EndFor \label{alg:predict--scoring-end}
  \State $\rhd$ Merging phase \label{alg:predict--merging-begin}
  \State $P \gets \{\}$ \textbf{;} $P_h \gets \{\}$
  \State $sort(P_t)$ \textbf{on each thread}
  \ForAll{$t$ in threads}
    \If{$|P_t| \neq 0$} $P_h.push(\{t, P_t.back().score\})$
    \EndIf
  \EndFor
  \State $P_h.makeMaxHeapByScore()$ 
  \While{$|P_h| \neq 0$ \textbf{and} $|P| < N_P$} \label{alg:predict--merge-loop-begin}
    \State $t \gets P_h.popHeap().t$
    \State $P.push(P_t.back())$
    \State $P_t.pop()$
    \If{$|P_t| \neq 0$} $P_h.pushHeap(\{t, P_t.back().score\})$
    \EndIf
  \EndWhile \label{alg:predict--merging-end}
  \Return{$P$}
\EndFunction \label{alg:predict--end}

\Statex

\Function{scanEdges}{$H_t, G, a, b$} \label{alg:predict--scan-begin}
  \ForAll{$c \in G.out(b)$}
    \State $\rhd$ Skip reverse edges
    \If{$c \leq a$} \textbf{continue}
    \EndIf
    \If{$metric = AA$} $H_t[c] \gets H_t[c] + 1 / log(|\Gamma_b|)$
    \ElsIf{$metric = RA$} $H_t[c] \gets H_t[c] + 1 / |\Gamma_b|$
    \Else\ $H_t[c] \gets H_t[c] + 1$
    \EndIf
  \EndFor
\EndFunction \label{alg:predict--scan-end}
\end{algorithmic}
\end{algorithm}

\section{Evaluation}
\label{sec:evaluation}
\subsection{Experimental Setup}
\label{sec:setup}

\subsubsection{System used}

We use a server outfitted with two Intel Xeon Gold 6226R processors. Each processor comprises $16$ cores operating at $2.90$ GHz. Each core has a $1$ MB L1 cache, a $16$ MB L2 cache, and a shared L3 cache of $22$ MB. The system is set up with $376$ GB of system memory and has CentOS Stream 8 installed.

\subsubsection{Configuration}

We employ 32-bit integers to represent vertex IDs and 32-bit floats for score computation. We utilize $32$ threads to match the system core count (unless specified otherwise). Compilation is carried out using GCC 8.5 and OpenMP 4.5.

\subsubsection{Dataset}

The graphs used in our experiments are given in Table \ref{tab:dataset}. These are sourced from the SuiteSparse Matrix Collection \cite{suite19}. In the graphs,\ignore{the} number of vertices vary from $3.07$ to $214$ million, and\ignore{the} number of edges vary from $25.4$ million to $3.80$ billion. We ensure that edges are undirected and weighted with a default of $1$.

\begin{table}[hbtp]
  \centering
  \caption{List of $13$ graphs obtained from the SuiteSparse Matrix Collection \cite{suite19}, with the directed graphs being marked with an asterisk. Here, $|V|$ is the number of vertices, $|E|$ is the number of edges (after adding reverse edges and removing self-loops), and $D_{avg}$ is the average degree.}
  \label{tab:dataset}
  \begin{tabular}{|c||c|c|c|}
    \toprule
    \textbf{Graph} &
    \textbf{\textbf{$|V|$}} &
    \textbf{\textbf{$|E|$}} &
    \textbf{\textbf{$D_{avg}$}} \\
    \midrule
    \multicolumn{4}{|c|}{\textbf{Web Graphs (LAW)}} \\ \hline
    indochina-2004$^*$ & 7.41M & 339M & 45.7 \\ \hline
    uk-2002$^*$ & 18.5M & 561M & 30.3 \\ \hline
    arabic-2005$^*$ & 22.7M & 1.20B & 52.8 \\ \hline
    uk-2005$^*$ & 39.5M & 1.71B & 43.4 \\ \hline
    webbase-2001$^*$ & 118M & 1.86B & 15.8 \\ \hline
    it-2004$^*$ & 41.3M & 2.17B & 52.6 \\ \hline
    sk-2005$^*$ & 50.6M & 3.78B & 74.7 \\ \hline
    \multicolumn{4}{|c|}{\textbf{Social Networks (SNAP)}} \\ \hline
    com-LiveJournal & 4.00M & 69.4M & 17.3 \\ \hline
    com-Orkut & 3.07M & 234M & 76.3 \\ \hline
    \multicolumn{4}{|c|}{\textbf{Road Networks (DIMACS10)}} \\ \hline
    asia\_osm & 12.0M & 25.4M & 2.1 \\ \hline
    europe\_osm & 50.9M & 108M & 2.1 \\ \hline
    \multicolumn{4}{|c|}{\textbf{Protein k-mer Graphs (GenBank)}} \\ \hline
    kmer\_A2a & 171M & 361M & 2.1 \\ \hline
    kmer\_V1r & 214M & 465M & 2.2 \\ \hline
  \bottomrule
  \end{tabular}
\end{table}

\subsubsection{Generating Observed graph and Unobserved edges}
\label{sec:generate-batch}

To generate the observed graph $E^O$ and unobserved edges $E^U$, we randomly remove $10^{-2}|E|$ to $0.1|E|$ edges from each graph in the dataset. The removed edges constitute the unobserved edges $E^U$, with the endpoints chosen uniformly at random \cite{zhou2021progresses}. The number of unobserved edges $|E^U|$ is typically set at $10\%$ of the total links in $E$, i.e., $0.1|E|$ based on empirical findings, ensuring statistically robust results without significantly altering the network's structure \cite{lu2015toward}.

\subsubsection{Expected precision of Random guess}

We now discuss the expected precision of a random guess (instead of using a link prediction algorithm). For an observed graph $G(V, E)$, with $N$ vertices and $M$ edges, there are $N(N-1)/2 - M$ possible links, and thus the expected precision for predicting $|P|$ edges is $1 / {}_{{N(N-1)/2 - M}} C_{|P|}$. This is an incredibly small number. For instance, on a graph with $10$ vertices and $100$ edges, correctly predicting all $|P| = 10$ has a probability of $3\times10^{-11}$.

\subsubsection{Measuring Prediction quality}

As previously stated, we rely on the F1 score, the harmonic mean of precision and recall, to evaluate link prediction performance \cite{lu2015toward}. This choice is made due to concerns that AUC may inaccurately favor algorithms ranking many negatives at the bottom \cite{zhou2021progresses, yang2015evaluating, lichtnwalter2012link}.

\subsubsection{Missing links in the original graphs}

There might be missing links in the original dataset graphs, and link prediction algorithms could attempt to predict them, although we currently lack a method to verify this. While working with temporal graphs could address this issue, the available temporal graphs are not sufficiently large. As a result, our current focus is on static graphs in the dataset, while generating observed graphs and unobserved edges. The exploration of temporal graphs is planned for future work.

\subsubsection{Heterogeneity of the graphs}

Real-world graphs are often heterogeneous, with diverse linking patterns throughout the graph. Consequently, using a single link prediction method may not be ideal. Instead, employing different link prediction methods on distinct regions of the graph may be more suitable. However, our belief is that a specific linking pattern dominates the graph, allowing us to identify an overall suitable link prediction method.

\subsection{Comparative Performance Evaluation}

\begin{figure*}[hbtp]
  \centering
  \subfigure[Runtime in seconds (logarithmic scale) for link prediction using the best similarity measure, with \textit{IBase} and \textit{DLH} approaches]{
    \label{fig:input-large--runtime}
    \includegraphics[width=0.98\linewidth]{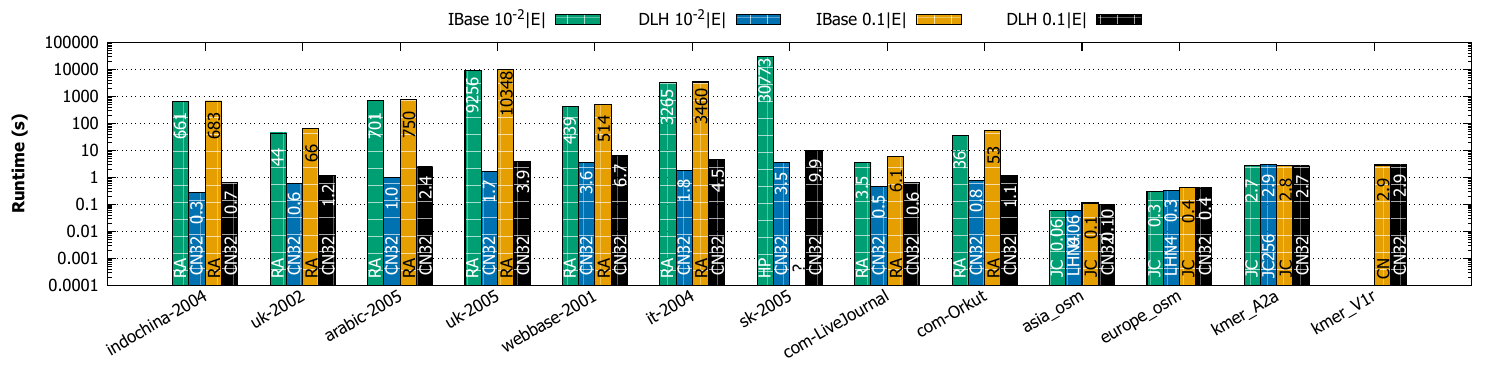}
  }
  \subfigure[Speedup (logarithmic scale) for link prediction with the best similarity measure of \textit{DLH} approach, compared to \textit{IBase} approach]{
    \label{fig:input-large--speedup}
    \includegraphics[width=0.98\linewidth]{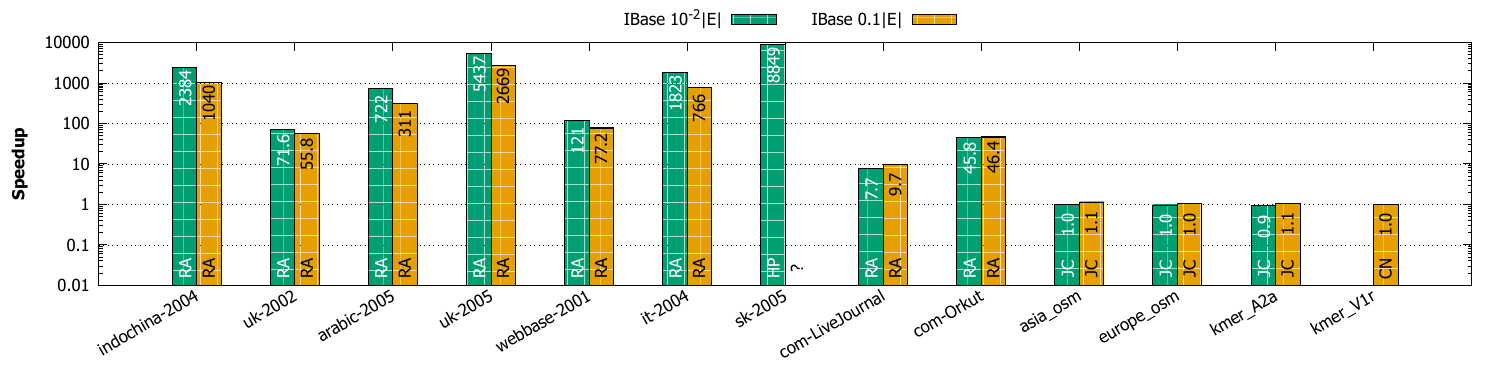}
  }
  \subfigure[F1 score of predicted links (logarithmic scale), for link prediction using the best similarity measure, with \textit{IBase} and \textit{DLH} approaches]{
    \label{fig:input-large--f1score}
    \includegraphics[width=0.98\linewidth]{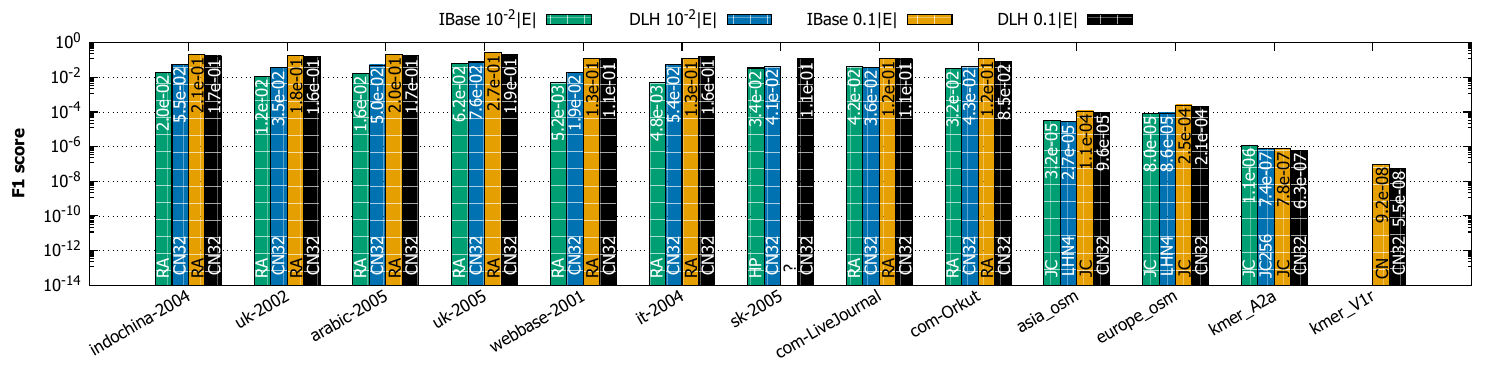}
  } \\[-2ex]
  \caption{Runtime in seconds (log-scale), speedup (log-scale), and F1 score of predicted links (log-scale), for link prediction with the \textit{Improved Baseline (IBase)} approach and our approach of \textit{Disregarding Large Hubs (DLH)}, using the best similarity measure, when attempting to predict $10^{-2}|E|$ to $0.1|E|$ unobserved edges $E^U$, for each graph\ignore{in the dataset}. For each similarity measure outlined in Section \ref{sec:metrics}, we attempt only the best hub limit $L_H$ parameter setting obtained in Section \ref{sec:select-limit} (for the \textit{DLH} approach), and then select the best among them, considering both the F1 score and runtime. Note that the numerical suffix added to the acronym of each link prediction method, with the \textit{DLH} approach, indicates the hub limit $L_H$ parameter setting.}
  \label{fig:input-large}
\end{figure*}

In this section, we compare the performance of the DLH approach, which disregards large hubs, with the IBase approach (which does not). We conduct this comparison for observed graphs $E^O$ based on each graph in the dataset, with $10^{-2}|E|$ to $0.1|E|$ unobserved edges. This involves removing $10^{-2}|E|$ to $0.1|E|$ edges, with endpoints chosen uniformly at random (as explained in Section \ref{sec:generate-batch}). We predict the same number of edges with both approaches, i.e., $N_P = 10^{-2}|E|$ or $0.1|E|$. For each observed graph, we predict links using the IBase approach and the DLH approach with a suitable hub limit $L_H$ identified in Section \ref{sec:select-limit}. In Figures \ref{fig:input-large--runtime}, \ref{fig:input-large--speedup}, and \ref{fig:input-large--f1score}, we plot the runtimes, speedups, and F1 scores, respectively, for only the best approach for each graph (considering both F1 score and runtime). In the figures, the labels indicate the abbreviations of the similarity metric used, followed by the value of the hub limit $L_H$ parameter setting (e.g., the label $CN32$ stands for the Common Neighbors (CN) similarity metric, with a hub limit $L_H$ of $32$, where first order neighbors with a degree greater than $32$ are avoided). It's worth noting that a hub limit $L_H$ of $\infty$ essentially represents the IBase approach. Note that the IBase approach crashed on the \textit{sk-2005} graph with $0.1|E|$ unobserved edges due to out-of-memory issue, and thus, these plots are not shown.

\begin{figure}[hbtp]
  \centering
  \includegraphics[width=0.98\linewidth]{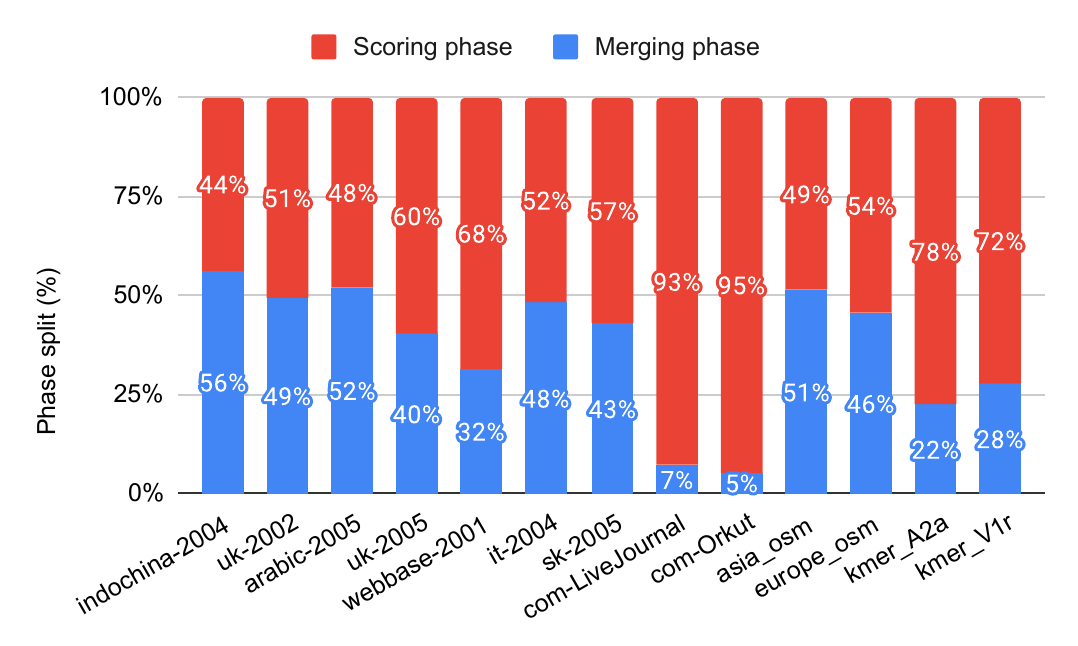} \\[-2ex]
  \caption{Overall phase split of the \textit{DLH} approach, when predicting $0.1|E|$ links with similarity measures given in Section \ref{sec:metrics}, while choosing suitable hub limit $L_H$ setting for each similarity measure as given in Section \ref{sec:select-limit}.}
  \label{fig:phase-split}
\end{figure}

\begin{figure}[hbtp]
  \centering
  \includegraphics[width=0.98\linewidth]{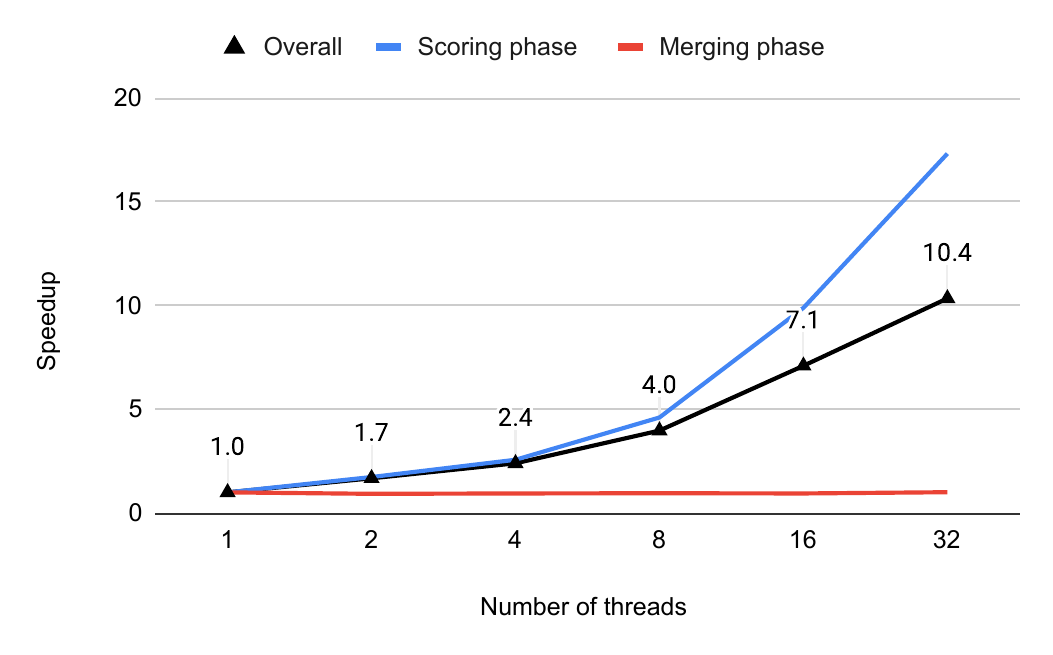} \\[-2ex]
  \caption{Overall speedup of our approach of \textit{Disregarding Large Hubs (DLH)} for link prediction, and its phases (obtaining edges with top-k scores per thread, and merging scores from each thread into a common scoreboard), with $10^{-2}|E|$ unobserved edges, with increasing number of threads (in multiples of 2). Increasing the number of threads causes work in the merging phase to increase,\ignore{thus} leading to a poor speedup.}
  \label{fig:strong-scaling}
\end{figure}

As seen in Figure \ref{fig:input-large}, the DLH approach is, on average, over $1622\times$ and $415\times$ faster than the IBase approach with $10^{-2}|E|$ and $0.1|E|$ unobserved edges, respectively. It achieves this speedup while predicting links with an average F1 score that is $80\%$ higher and $13\%$ lower, respectively --- meaning similar F1 scores without being too low or high. Notably, on the \textit{sk-2005} graph with $0.1|E|$ edges removed, DLH achieves a link prediction rate of $38.1M$ edges/s.

Furthermore, we observe that link prediction with the RA metric excels, in terms of both F1 score and runtime, on web graphs and social networks when using the IBase approach. Meanwhile, link prediction with the JC similarity metric outperforms others on road networks and protein k-mer graphs. With the DLH approach and $10^{-2}|E|$ unobserved edges, link prediction with the CN metric (hub limit $L_H$ of $32$) excels on web graphs and social networks. For road networks, link prediction with the LHN metric (hub limit $L_H$ of $4$) performs the best, and for protein k-mer graphs, link prediction with the JC metric (hub limit $L_H$ of $256$) is optimal. However, with $0.1|E|$ unobserved edges, link prediction with the CN metric (hub limit $L_H$ of $32$) proves to be the best across all graphs. Figures \ref{fig:standard2}, \ref{fig:standard1}, \ref{fig:pruned2}, \ref{fig:pruned1} show the runtimes and F1 scores for link prediction with all similarity measures. Notably, in Figures \ref{fig:pruned2} and \ref{fig:pruned1}, the DLH approach using the AA and RA metrics performs similarly to the CN metric, but with longer runtimes.

Next, we note that the IBase approach achieves an average F1 score of $1.8\times10^{-2}$ and $1.1\times10^{-1}$ when predicting $10^{-2}|E|$ and $0.1|E|$ edges, respectively. In comparison, the DLH approach achieves F1 scores, averaging $3.2\times10^{-2}$ and $9.8\times10^{-2}$, respectively. Additionally, we observe that the F1 score tends to be higher on web graphs and social networks but significantly lower on road networks and protein k-mer graphs. This discrepancy is likely due to the average degree of the graphs, as local/neighborhood-based link prediction methods rely on the neighborhood of vertices up to a distance of $2$. Graphs with lower average degrees provide less information to such methods for predicting edges. While these F1 scores may seem low (compared to the highest possible F1 score of $1$), it's important to consider that they are significantly higher than random predictions. Although machine learning-based approaches might achieve higher F1 scores, neighborhood-based similarity measures excel in computational efficiency (both in terms of runtime and space) and interpretability, as mentioned earlier.

\ignore{Why the choice of best method appears to change with batch size?}
\ignore{Why do these methods perform well?}
\ignore{What we observe on our dataset? What we observe on smaller graphs?}

\subsection{Performance Analysis}

In this section, we examine the phase split of the DLH approach to identify further optimization opportunities. To do this, we generate an observed graph $E^O$ with $0.1|E|$ unobserved edges $E^U$ for each dataset graph using random division, as detailed in Section \ref{sec:generate-batch}. For each observed graph, we predict $0.1|E|$ links using the DLH approach and the similarity measures outlined in Section \ref{sec:metrics}, utilizing the appropriate hub limit $L_H$ value, as per Section \ref{sec:select-limit}.

Figure \ref{fig:phase-split} shows that DLH, spends a majority of its runtime, $63\%$ on average, in the scoring phase. This is particularly notable in social networks with a high average degree. However, a substantial amount of time is still spent on the merging phase, which involves combining predicted edges with top-$k$ scores from each thread into a global top-$k$ list of predicted edges. The sequential nature of the merging phase likely contributes to this runtime, and addressing this aspect is a potential focus for future work.

\subsection{Strong Scaling}

In the final analysis, we evaluate the strong scaling performance of our DLH approach, where we perform link prediction while disregarding large hubs, i.e., first order neighbors with high degree. The assessment involves varying the number of threads from $1$ to $32$ in multiples of $2$ for each input graph. We measure the average time taken to predict $0.1|E|$ links using similarity measures defined in Section \ref{sec:metrics}, incorporating the best hub limit $L_H$ setting identified in Section \ref{sec:select-limit}. Figure \ref{fig:strong-scaling} presents the results, illustrating not only the overall scaling performance but also the scaling of the two phases of each link prediction method: identifying edges with top-$k$ scores in each thread (\textit{scoring phase}) and combining scores across threads to obtain the global top-$k$ edges (\textit{merging phase}).

With $32$ threads, the DLH approach achieves an overall speedup of $10.4\times$ compared to sequential execution, indicating a performance increase of $1.6\times$ for every doubling of threads. The scalability is limited, as the cost of the merging phase increases with an increase in the number of threads, and because the merging is performed with sequential execution. In fact, at $32$ threads, the merging phase obtains no speedup of $1.0\times$, while the scoring phase achieves a speedup of $17.3\times$.

\section{Conclusion}
\label{sec:conclusion}
Link prediction aims to predict missing or future connections between nodes based on existing interactions and network structure. Similarity measures are widely used for their simplicity, interpretability, and efficiency, and are often combined with methods like ensemble learning. Evaluating these algorithms on large networks is crucial, but many studies focus on small to medium-scale graphs. To address this, we present a parallel approach, \textit{IBase}, that efficiently finds common neighbors and handles large graphs by tracking top-$k$ edges per thread. We also introduce a novel heuristic that disregards large hubs (\textit{DLH}). It is based on the notion that low-degree nodes contribute significant similarity among neighbors, in contrast to high-degree nodes (interestingly, this is similar to the idea behind the AA and RA similarity metrics).

Our results show that the DLH approach is, on average, over $1622\times$ and $415\times$ faster than the IBase approach with $10^{-2}|E|$ and $0.1|E|$ unobserved edges, respectively, while maintaining similar F1 scores ($80\%$ higher and $13\%$ lower, respectively). On the \textit{sk-2005} graph with $0.1|E|$ edges removed, DLH predicts links at a rate of $38.1M$ edges/s. The RA metric performs best on web graphs and social networks using the IBase approach, while the JC metric excels on road networks and protein k-mer graphs. With DLH and $10^{-2}|E|$, CN (hub limit $L_H = 32$) works best for web graphs and social networks, LHN ($L_H = 4$) for road networks, and JC ($L_H = 256$) for protein k-mer graphs. With $0.1|E|$, CN ($L_H = 32$) is optimal across all graphs. DLH consistently outperforms IBase, especially on web graphs and social networks. For DLH with $0.1|E|$ edges, $63\%$ of the runtime is spent on scoring, particularly in social networks, with significant time also spent on merging, hinting at optimization opportunities. DLH achieves a $10.4\times$ speedup with $32$ threads, but scalability is limited by the merging phase.
Future work may focus on optimizing quasi-local and global similarity methods and extending our techniques to higher-order neighbors.
\begin{acks}
I would like to thank Prof. Kishore Kothapalli, Prof. K. Swarupa Rani, Ashwitha Gatadi, and Balavarun Pedapudi for their support.
\end{acks}

\bibliographystyle{ACM-Reference-Format}
\bibliography{main}

\clearpage
\appendix
\begin{figure*}[hbtp]
  \centering
  \subfigure[Speedup of \textit{DLH} approach with different hub limits $L_H$]{
    \label{fig:adjust-overall--runtime}
    \includegraphics[width=0.48\linewidth]{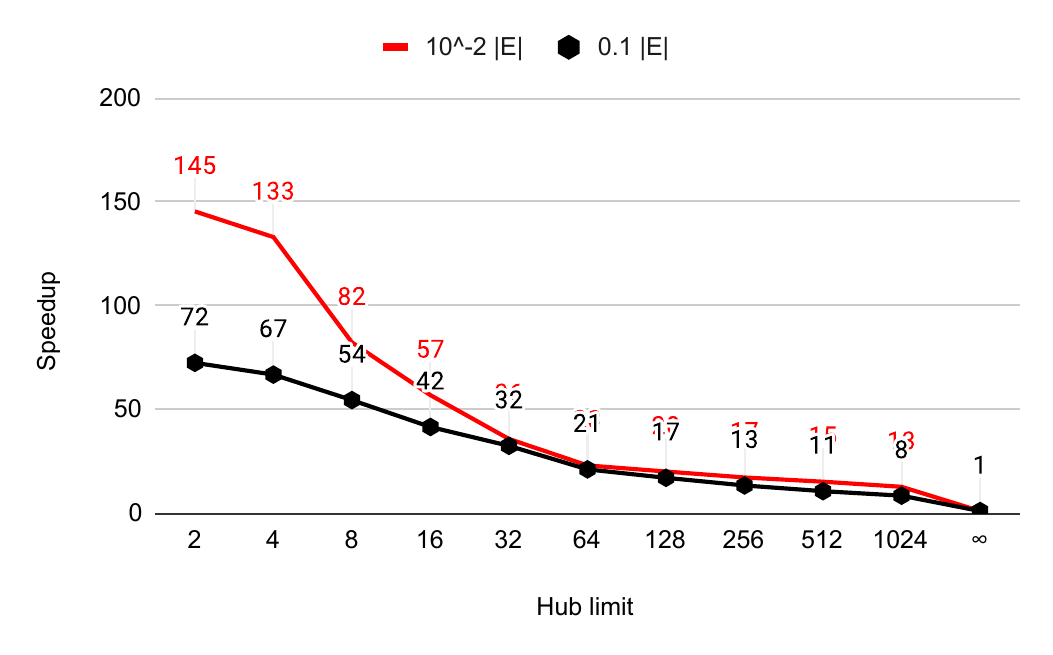}
  }
  \subfigure[F1 score of predicted links (logarithmic scale), with different hub limits $L_H$]{
    \label{fig:adjust-overall--precision}
    \includegraphics[width=0.48\linewidth]{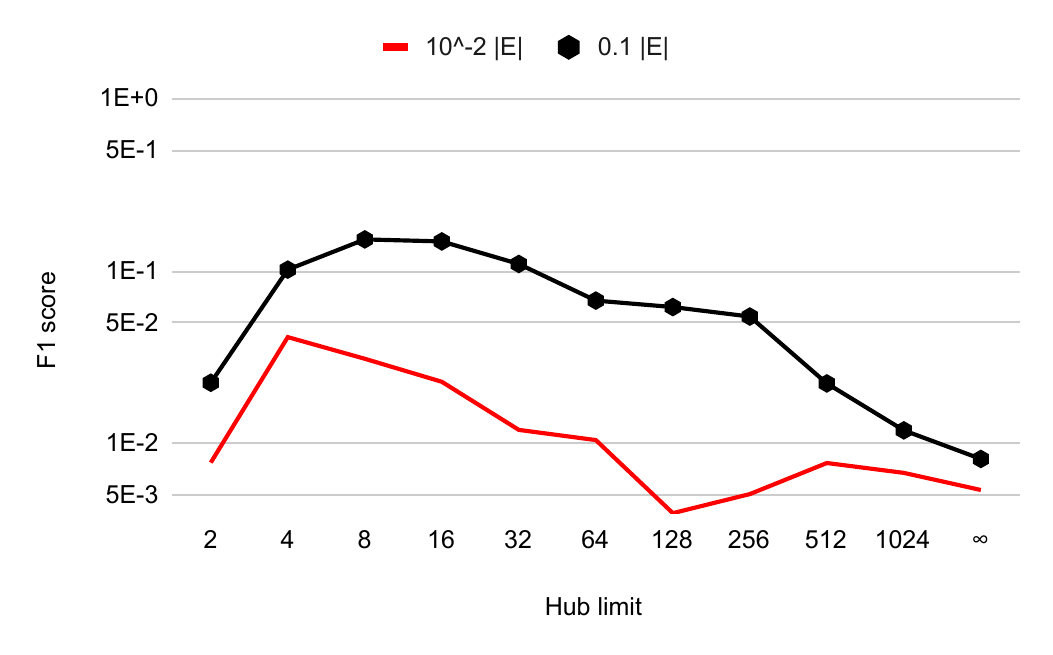}
  } \\[0ex]
  \caption{Overall impact of adjusting the hub limit $L_H$ from $2$ to $1024$ (in multiples of $2$), and to $\infty$, on the speedup and F1 score of predicted links (log scale), of neighbor-based link prediction methods, with the number of unobserved edges $E^U$ of $10^{-2}|E|$ and $0.1|E|$. Speedup is measured with respect to hub limit $L_H$ of $\infty$, i.e., the \textit{Improved Baseline (IBase)} approach.\ignore{, using geometric mean of runtimes for link prediction using the DLH approach using all similarity scores given in Section \ref{sec:metrics}; while overall F1 score is obtained by taking the average.}}
  \label{fig:adjust-overall}
\end{figure*}

\begin{figure*}[hbtp]
  \centering
  \subfigure[Runtime in seconds (logarithmic scale) for link prediction using various similarity measures, with \textit{IBase} approach]{
    \label{fig:standard2--runtime}
    \includegraphics[width=0.98\linewidth]{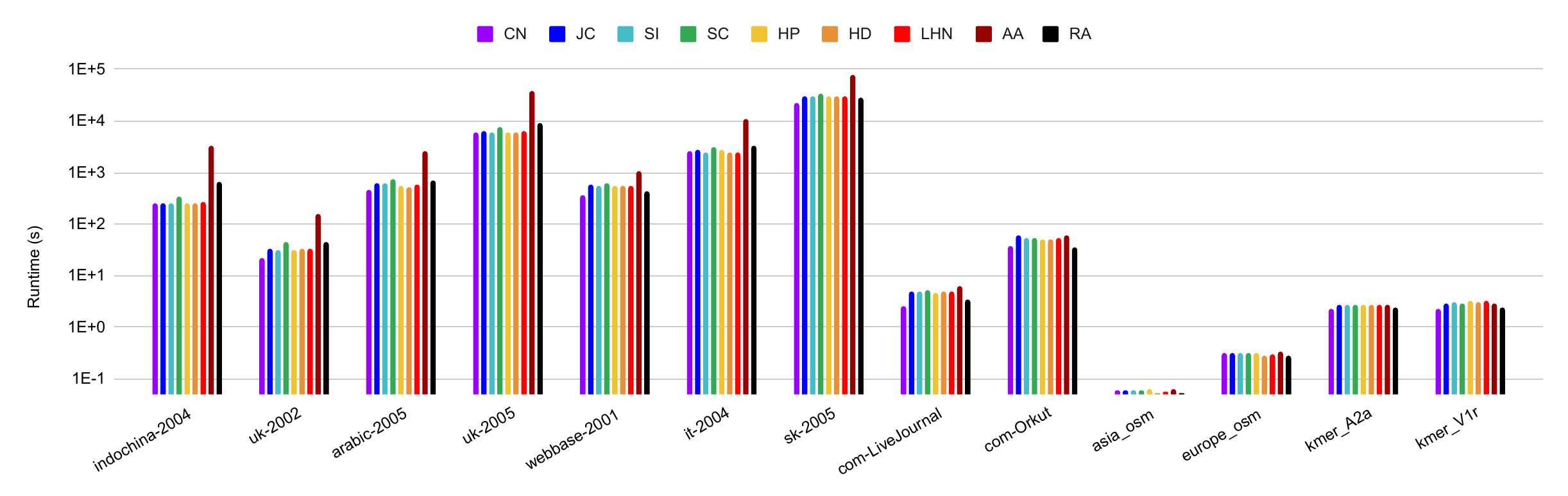}
  }
  \subfigure[F1 score of predicted links (logarithmic scale) for link prediction using various similarity measures, with \textit{IBase} approach]{
    \label{fig:standard2--f1score}
    \includegraphics[width=0.98\linewidth]{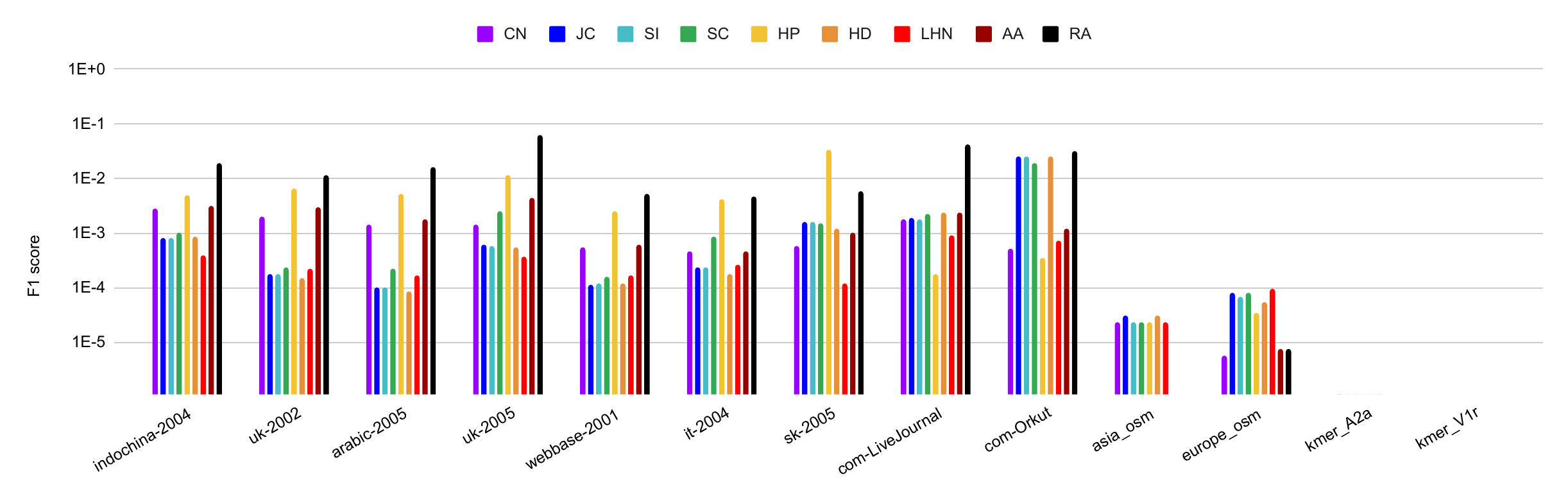}
  } \\[-2ex]
  \caption{Runtime (log-scale) and F1 score (log-scale) for link prediction using various similarity measures with \textit{Improved Baseline (IBase)} approach, when attempting to predict $10^{-2}|E|$ unobserved edges $E^U$ for each graph\ignore{in the dataset}.}
  \label{fig:standard2}
\end{figure*}

\begin{figure*}[hbtp]
  \centering
  \subfigure[Runtime in seconds (logarithmic scale) for link prediction using various similarity measures, with \textit{IBase} approach]{
    \label{fig:standard1--runtime}
    \includegraphics[width=0.98\linewidth]{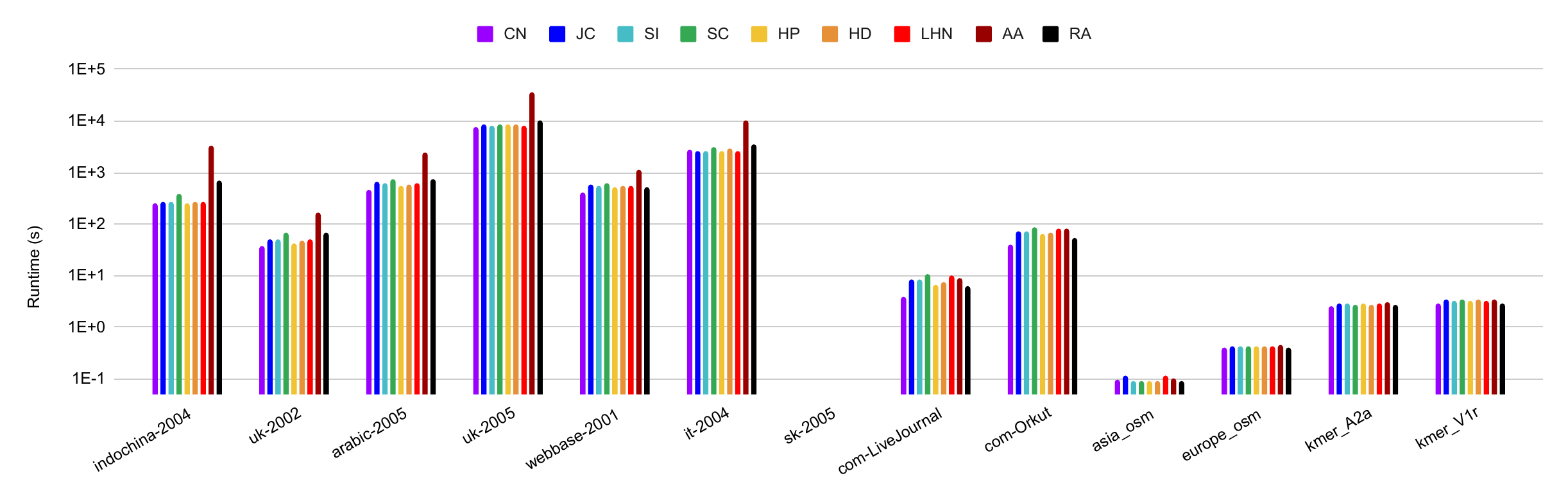}
  }
  \subfigure[F1 score of predicted links (logarithmic scale) for link prediction using various similarity measures, with \textit{IBase} approach]{
    \label{fig:standard1--f1score}
    \includegraphics[width=0.98\linewidth]{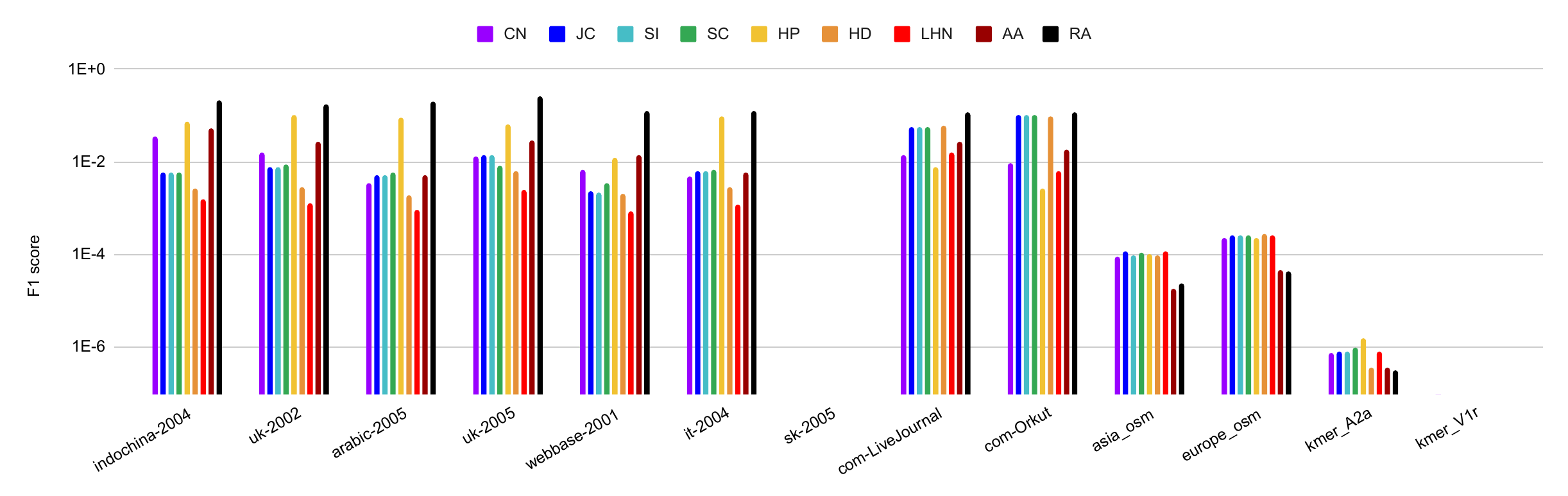}
  } \\[-2ex]
  \caption{Runtime (log-scale) and F1 score (log-scale) for link prediction using various similarity measures with \textit{Improved Baseline (IBase)} approach, when attempting to predict $0.1|E|$ unobserved edges $E^U$ for each graph\ignore{in the dataset}.}
  \label{fig:standard1}
\end{figure*}

\begin{figure*}[hbtp]
  \centering
  \subfigure[Runtime in seconds (logarithmic scale) for link prediction using various similarity measures, with \textit{DLH} approach]{
    \label{fig:pruned2--runtime}
    \includegraphics[width=0.98\linewidth]{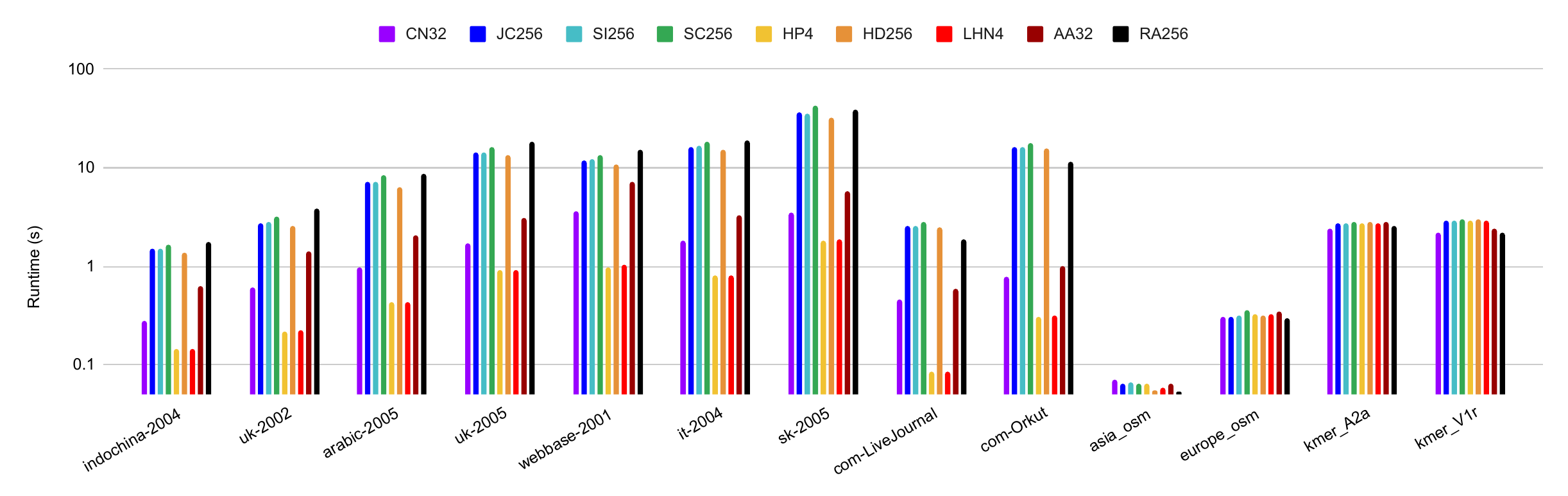}
  }
  \subfigure[F1 score of predicted links (logarithmic scale) for link prediction using various similarity measures, with \textit{DLH} approach]{
    \label{fig:pruned2--f1score}
    \includegraphics[width=0.98\linewidth]{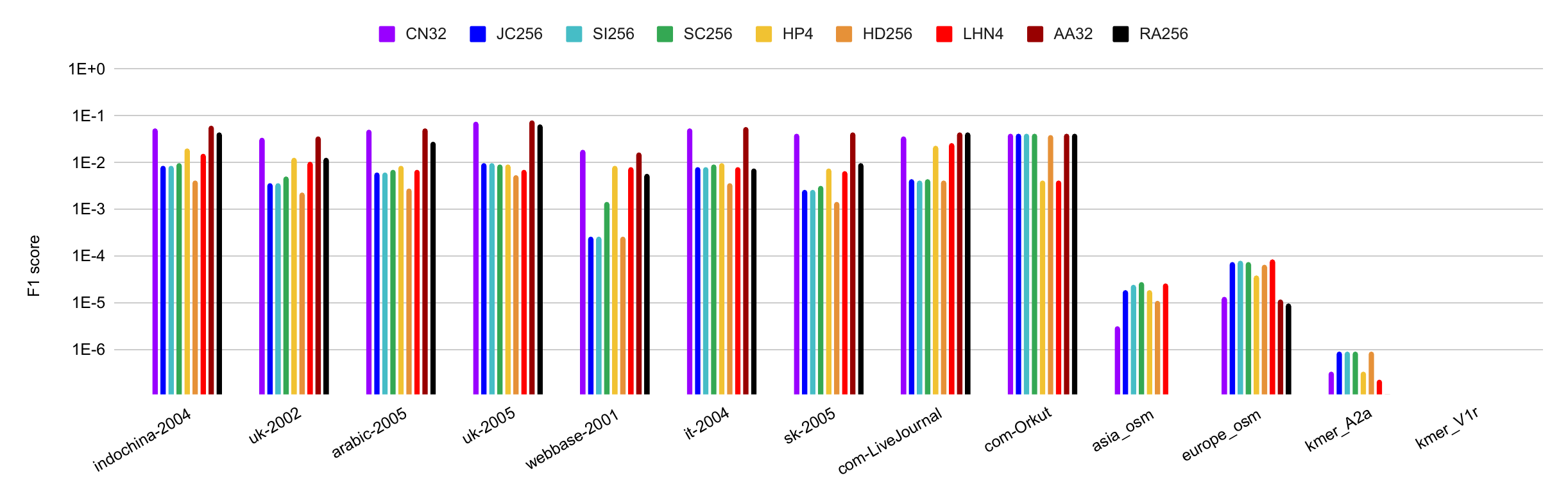}
  } \\[-2ex]
  \caption{Runtime (log-scale) and F1 score (log-scale) for link prediction using various similarity measures with our approach of \textit{Disregarding Large Hubs (DLH)}, when attempting to predict $10^{-2}|E|$ unobserved edges $E^U$ for each graph in the dataset. For each similarity measure outlined in Section \ref{sec:metrics}, the best hub limit $L_H$ parameter setting obtained in Section \ref{sec:select-limit} is used, indicated by a numerical suffix added to each link prediction method acronym.}
  \label{fig:pruned2}
\end{figure*}

\begin{figure*}[hbtp]
  \centering
  \subfigure[Runtime in seconds (logarithmic scale) for link prediction using various similarity measures, with \textit{DLH} approach]{
    \label{fig:pruned1--runtime}
    \includegraphics[width=0.98\linewidth]{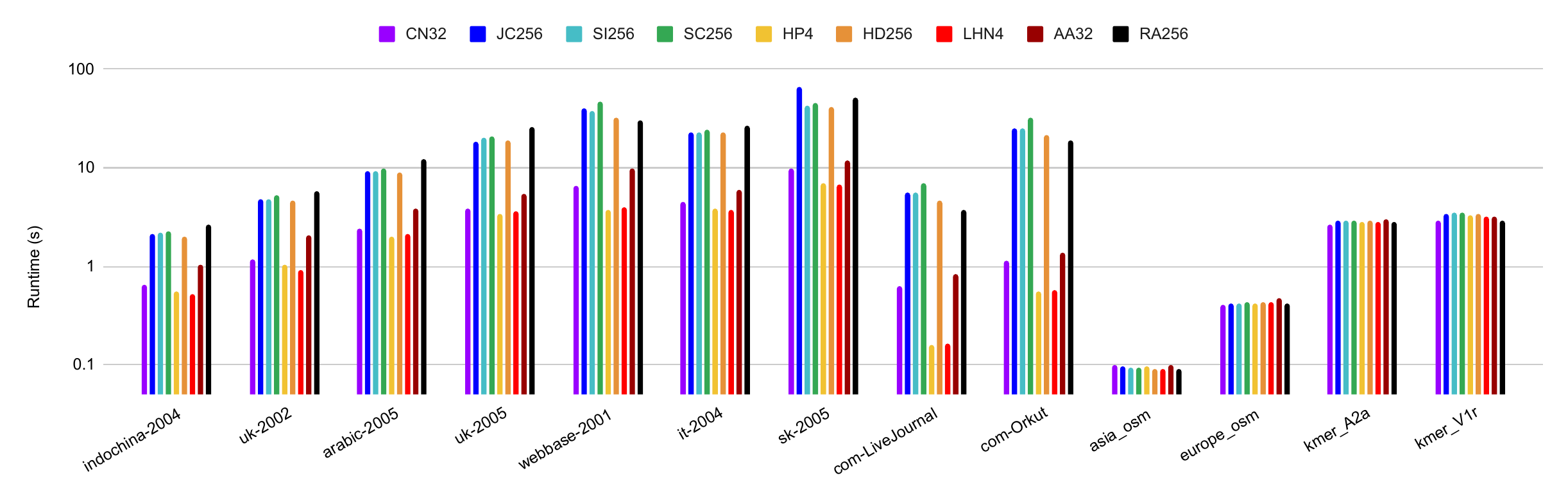}
  }
  \subfigure[F1 score of predicted links (logarithmic scale) for link prediction using various similarity measures, with \textit{DLH} approach]{
    \label{fig:pruned1--f1score}
    \includegraphics[width=0.98\linewidth]{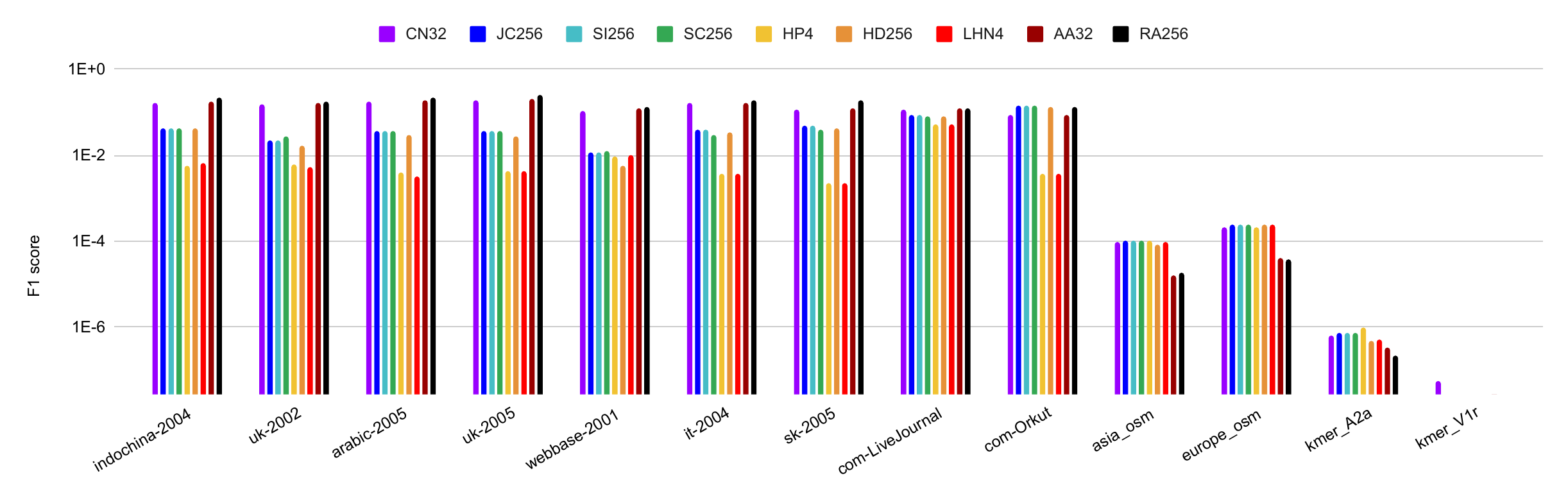}
  } \\[-2ex]
  \caption{Runtime (log-scale) and F1 score (log-scale) for link prediction using various similarity measures with our approach of \textit{Disregarding Large Hubs (DLH)}, when attempting to predict $0.1|E|$ unobserved edges $E^U$ for each graph in the dataset. For each similarity measure outlined in Section \ref{sec:metrics}, the best hub limit $L_H$ parameter setting obtained in Section \ref{sec:select-limit} is used, indicated by a numerical suffix added to each link prediction method acronym.}
  \label{fig:pruned1}
\end{figure*}

\end{document}